\documentclass[%
reprint,
groupedaddress,
amsmath,amssymb,
prb,
floatfix,
superscriptaddress
]{revtex4-1}

\usepackage{amsmath}
\usepackage{bm}
\usepackage{amsfonts}
\usepackage{graphicx}
\usepackage{amssymb}
\usepackage{sidecap}
\usepackage[caption=false]{subfig}
\usepackage{hyperref}
\usepackage{cleveref}
\usepackage{siunitx}
\usepackage[T1]{fontenc}
\usepackage{libertine}

\renewcommand{\vec}[1]{\bm{#1}}
\newcommand{\mat}[1]{\mathbf{#1}}
\newcommand{\mush}[0]{\textsc{MultiShifter}}

\begin{document}

\title{\mush{}: software to generate structural models of extended two-dimensional defects in 3D and 2D crystals}
\author{Jon Gabriel Goiri}
\affiliation{Materials Department, University of California Santa Barbara}
\author{Anton Van der Ven}
\email{avdv@ucsb.edu}
\affiliation{Materials Department, University of California Santa Barbara}
\date{\today}

\begin{abstract}
Extended defects in crystals, such as dislocations, stacking faults and grain boundaries, play a crucial role in determining a wide variety of materials properties. 
Extended defects can also lead to novel electronic properties in two-dimensional materials, as demonstrated by recent discoveries of emergent electronic phenomena in twisted graphene bilayers. 
This paper describes several approaches to construct crystallographic models of two-dimensional extended defects in crystals for first-principles electronic structure calculations, including (i) crystallographic models to parameterize generalized cohesive zone models for fracture studies and meso-scale models of dislocations and (ii) crystallographic models of twisted bilayers. 
The approaches are implemented in an open source software package called \mush{}.

\end{abstract}

\maketitle
\section{Introduction}\label{sec:intro}
Extended defects play an important role in many materials applications, affecting electronic, functional and mechanical properties. 
The simplest two-dimensional extended defect is a surface created upon cleaving two halves of a crystal \cite{dross2007stress,sweet2016controlled,shim2018controlled}. 
The two halves of a crystal can also be rigidly sheared relative to each other as occurs in many layered battery materials that undergo stacking sequence phase transformations upon intercalation \cite{radin2017role,kaufman2019understanding,van2020rechargeable}.
Plastic deformation of crystals is mediated by the passage of dislocations that glide along a slip plane. While a dislocation is an extended one-dimensional defect, it often dissociates into a pair of partial dislocations that bound an extended two-dimensional stacking fault \cite{hull2011introduction} . 

Two-dimensional extended defects are also present in heterostructures of 2D materials \cite{mounet2018two}. 
2D materials exhibit unique electronic properties that are absent in their three dimensional counterparts. 
Furthermore, new electronic behavior can emerge when a pair of two-dimensional building blocks are twisted relative to each other \cite{shallcross2008quantum,mele2010commensuration,shallcross2010electronic}, as was recently demonstrated for graphene \cite{cao2018correlated,cao2018unconventional}. 
The twisting of a pair of stacked two-dimensional materials by an angle $\theta$ around an axis that is perpendicular to the sheets breaks the underlying translational symmetry of the 2D building blocks. 
A new field of "twistronics" has emerged that seeks to understand and exploit the electronic properties that arise upon twisting a pair of two-dimensional materials \cite{carr2020electronic}.

Here we describe a generalized framework and accompanying software package, called \mush{}\cite{mushgithub}, to facilitate the study of the thermodynamic and electronic properties of periodic two-dimensional defects in crystalline solids from first principles. 
We introduce the concept of a generalized cohesive zone model that simultaneously encapsulates the energy of decohesion and rigid glide of two halves of a crystal. 
We then detail how super cells and crystallographic models can be constructed to accommodate a pair of two-dimensional slabs that have been twisted relative to each other by an angle $\theta$.
\mush{}\cite{mushgithub} automates the construction of crystallographic models of extended two-dimensional defects and of two-dimensional layered materials to enable (i) the study of surfaces, (ii) the calculation of cohesive zone models \cite{jarvis2001effects,friak2003ab,van2004thermodynamics,hayes2004universal,enrique2014solute,enrique2017traction,olsson2015role,olsson2016first,olsson2017intergranular} used as constitutive laws for fracture studies \cite{deshpande2002discrete, xie2006discrete,sills2013effect}, (iii) the calculation of generalized stacking fault energy surfaces needed as input for phase field \cite{koslowski2002phase,shen2003phase,shen2004incorporation,hunter2011influence,feng2018shearing} and Peierls-Nabarro \cite{bulatov1997semidiscrete,juan1996generalized,lu2000peierls,lu2000generalized,lu2001dislocation,lu2001hydrogen,lu2005peierls} models of dislocations and (iv) the construction of models of twist grain boundaries \cite{sutton1995} and twisted 2D materials \cite{carr2020electronic}.
The crystallographic models generated by \mush{} can then be fed into first-principles electronic structure codes to calculate a range of thermodynamic and electronic properties. 
\mush{} consists of C++ and Python routines and draws on crystallographic libraries of the CASM software package \cite{thomas2013finite,puchala2013thermodynamics,van2018first}.

\section{Mathematical descriptions}\label{sec:math}

\subsection{Degrees of freedom}
\begin{figure}
\centering
\subfloat[]{
    \includegraphics[width=0.87\linewidth]{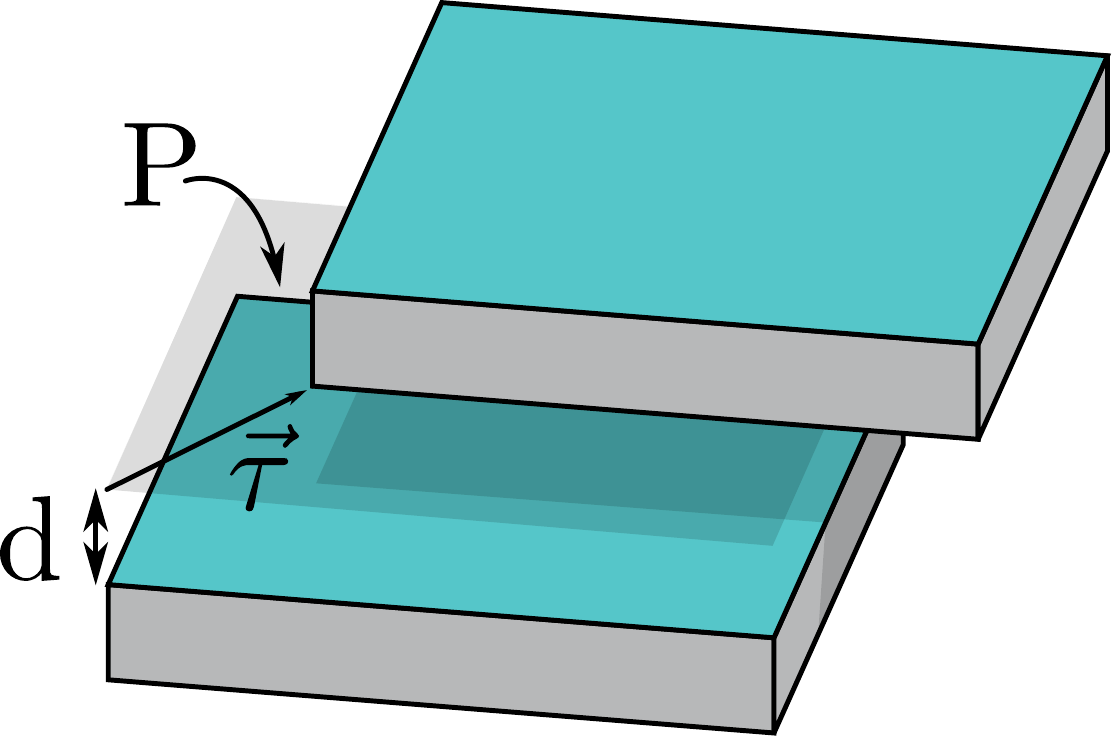}
    \label{fig:shiftschematic}
}
\hfill
\subfloat[]{
    \includegraphics[width=0.7\linewidth]{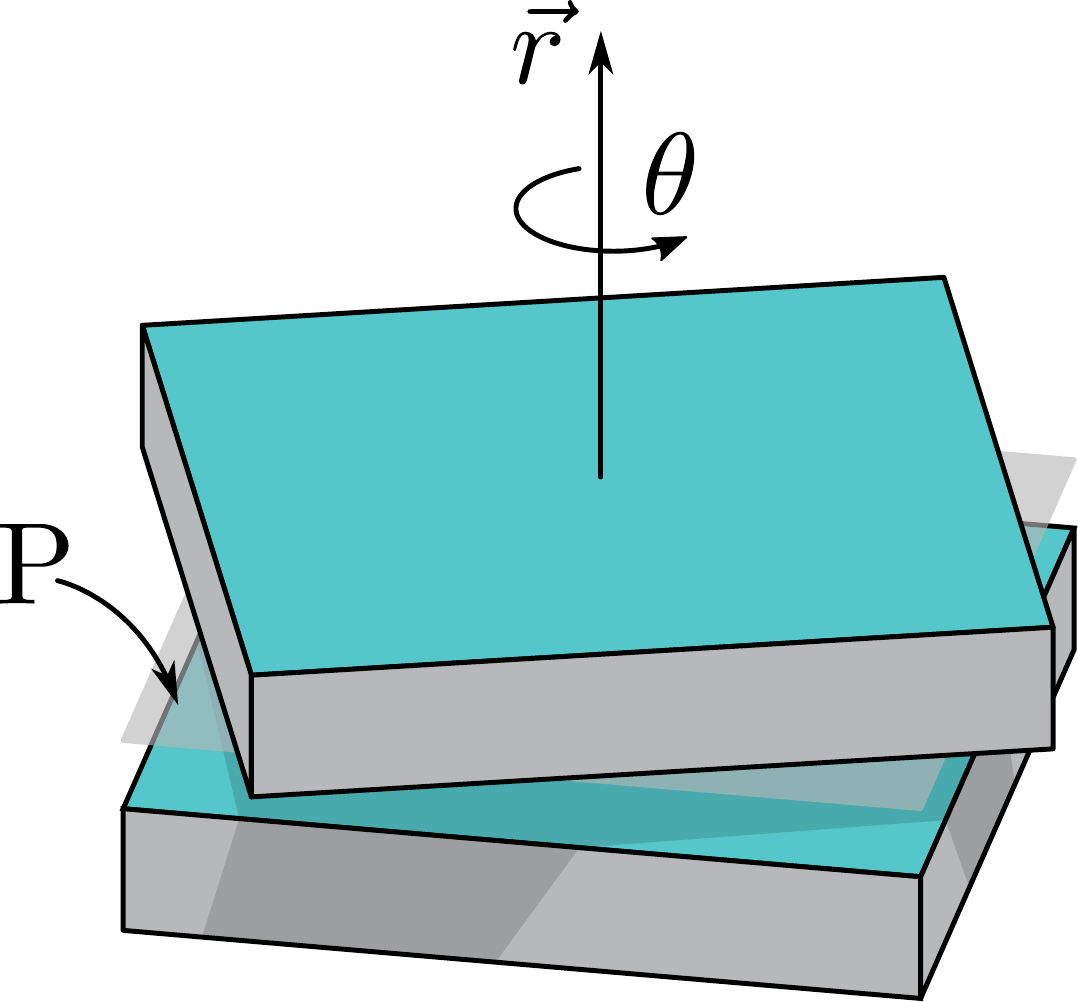}
    \label{fig:twistschematic}
}
\caption{(a) Two halves of a crystal separated by a distance $d$ can glide by a vector $\vec{\tau}$ parallel to the plane P.
(b) The two halves of a crystal can also be twisted by an angle $\theta$ around a twist axis $\vec{r}$ perpendicular to the plane $P$.}
\label{fig:schematics} 
\end{figure}

We consider displacements of two half crystals relative to a particular crystallographic plane {\bf P} as illustrated in \Cref{fig:schematics}. 
The half crystals could be two-dimensional materials such as a pair of graphene sheets or two-dimensional slabs of MoS$_2$. 
They could also be the bottom and top half of a macroscopic crystal that has a stacking fault or that is undergoing cleavage.
There are several ways in which the two half crystals can be displaced. 
As schematically illustrated in \Cref{fig:shiftschematic}, they may be separated by a distance $d$ along a direction perpendicular to the plane {\bf P} and they can be made to glide relative to each other by a translation vector $\vec{\tau}$ parallel to the plane {\bf P}. One half can also be twisted relative to the other by an angle $\theta$ around a rotation axis $\vec{r}$ that is perpendicular to the plane {\bf P} as illustrated in \Cref{fig:twistschematic}.
It is rare that uniform deformations across an infinite plane {\it P} as illustrated in \Cref{fig:schematics} occur in actual materials processes.
Nevertheless, the energy and electronic structure associated with such idealized deformations are crucial ingredients for a wide variety of mesoscale models of plastic deformation and fracture \cite{deshpande2002discrete, xie2006discrete,sills2013effect,koslowski2002phase,shen2003phase,shen2004incorporation,hunter2011influence,feng2018shearing,lu2000peierls,lu2000generalized,lu2001dislocation,lu2001hydrogen,lu2005peierls} and help understand emergent electronic properties in two-dimensional materials \cite{carr2020electronic}. 

\subsection{Mathematical expression for a generalized cohesive zone model}

A generalized cohesive zone model describes the energy of a bicrystal as the two half crystals are rigidly separated by $d$ and translated relative to each other by $\vec{\tau}$.
A convenient reference state for the energy scale is the bicrystal at infinite separation (i.e. $d\rightarrow{} \infty$). 
In this state, the energy of the bicrystal is independent of separation $d$ and translation $\vec{\tau}$. 
There are well-known and well tested functional forms for the dependence of the energy on separation \cite{rose1981universal,rose1983universal,rose1984universal,enrique2017traction,enrique2017decohesion}. 
A general relationship takes the form \cite{enrique2014solute,enrique2017traction}
\begin{equation}
    u_{cz}(d,\vec{\tau})=-2\gamma\left[ 1+\frac{\delta}{\lambda}+\sum_{n=2}^{n_{max}} \alpha_{n}\left(\frac{\delta}{\lambda}\right)^{n}\right]e^{-\delta/\lambda} 
    \label{eq:xuber}
\end{equation}
where $\delta = d-d_0$ measures the degree of separation relative to an equilibrium separation of $d_0$ and where the energy $u_{cz}$ is per unit area of the plane {\bf P}.
The dependence of the energy on the translation vector $\vec{\tau}$ can be built into \Cref{eq:xuber} by making the parameters $\gamma$, $\lambda$ and $\alpha_{n}$ each functions of the translation vector $\vec{\tau}$: i.e. $\gamma\left(\vec{\tau}\right)$, $\lambda\left(\vec{\tau}\right)$ and $\alpha_{n}\left(\vec{\tau}\right)$.
The equilibrium separation $d_{0}$ is also a function of $\vec{\tau}$ and corresponds to the minimum of $u_{cz}(\delta,\vec{\tau})$ as a function of $d$ at fixed $\vec{\tau}$.
The dependence of $u_{cz}\left(\delta,\vec{\tau}\right)$ on $\vec{\tau}$ at fixed $\delta$ or $d$ is periodic and has the same 2-dimensional periodicity as that of the crystallographic plane {\bf P}. 
This means that the parameters that appear in \Cref{eq:xuber} are also periodic functions of $\vec{\tau}$.

When all the $\alpha_{n}$ are set to zero, we recover the universal binding energy relation (UBER) that is able to describe the cohesive properties of metals with remarkable accuracy \cite{rose1981universal,rose1983universal,enrique2017decohesion}. 
The additional parameters $\alpha_{n}$ are necessary to capture deviations from the UBER form due to deviations from purely metallic bonding \cite{enrique2017traction}.

The coefficient $2\gamma$ is related to the energy of cleaving a crystal into two bicrystals
\begin{equation}
    2\gamma\left(\vec{\tau}\right)=\frac{E_{cleaved}-E_{bulk}\left(\vec{\tau}\right)}{A_{surface}}
    \label{eq:surface_energy}
\end{equation}
where $E_{bulk}\left(\vec{\tau}\right)$ is the energy of the bicrystal at separation $d$=$d_0\left(\vec{\tau}\right)$ and translation $\vec{\tau}$, $E_{cleaved}$ is the energy of the bicrystals at infinite separation and $A_{surface}$ is the area of the exposed surfaces. 
$2\gamma\left(\vec{\tau}\right)$ corresponds to the minimum of $u_{cz}\left(\delta,\vec{\tau}\right)$ at each $\vec{\tau}$ with respect to interslab separation $\delta$ and is referred to as the generalized stacking fault energy (GSFE), also known as the $\gamma$ surface. 
The GSFE is an essential ingredient of mesoscale simulation techniques such as phase field models \cite{koslowski2002phase,shen2003phase,shen2004incorporation,hunter2011influence,feng2018shearing} and Peierls-Nabarro \cite{lu2000peierls,lu2000generalized,lu2001dislocation,lu2001hydrogen,lu2005peierls} models of dislocations.

When cleaving a single crystal across a plane {\bf P} (as opposed to a bicrystal consisting of two different materials), it is common to set the origin of translations $\vec{\tau}$ at the shift coinciding with a perfect crystal (i.e. no stacking fault). 
In that case $\gamma_{0} = \gamma\left(\vec{\tau}=0\right)$ becomes equal to the surface energy for the crystallographic plane {\bf P} in the absence of any surface reconstructions \cite{thomas2010systematic}.
The generalized cohesive zone model can serve as a constitutive law to describe the response of a solid ahead of a transgranular crack tip \cite{deshpande2002discrete, xie2006discrete,sills2013effect}.
The elastic constant, $C$, along the direction of separation is a function of the parameters of \Cref{eq:xuber} according to \cite{enrique2017traction}
\begin{equation}
    C=2d_{0}\frac{2\gamma}{\lambda^2}\left(\frac{1}{2}-\alpha_2\right)
    \label{eq:elastic_constant}
\end{equation}

Since the parameters, $d_0$, $\gamma$, $\lambda$ and $\alpha_{n}$ of \Cref{eq:xuber} are periodic functions of $\vec{\tau}$, they can each be expressed as a Fourier series. For example, the $\vec{\tau}$ dependence of $\gamma$ can be expressed as
\begin{equation}
    \gamma\left(\vec{\tau}\right)=\sum_{\vec{K}}\tilde{\gamma}_{\vec{K}}e^{-i\vec{K}\vec{\tau}}
    \label{eq:fourier}
\end{equation}
where the sum extends over $\vec{K}$ vectors of the two-dimensional reciprocal lattice of the two-dimensional unit cell of the crystallographic plane {\bf P}.
The expansion coefficients, $\tilde{\gamma}_{K}$, are the Fourier transform of $\gamma$.

\subsection{Extensions to account for twist}\label{sec:twist}

The twisting of two halves of a crystal or of a pair of two-dimensional materials will generally break any translational symmetry that may have existed before. 
It is only for a subset of special twist angles $\theta$ that a super cell translational symmetry is preserved \cite{shallcross2008quantum,shallcross2010electronic,mele2010commensuration,silva2020exploring}. 
The energy of the bicrystal will not only depend on the relative separation $d$ and translation $\vec{\tau}$, but also on the choice of rotation axis $\vec{r}$ and twist angle $\theta$ \cite{silva2020exploring}.
This dependence can be formulated generally as 
\begin{equation}
    u\left(d,\vec{\tau},\vec{r},\theta\right)= u_{cz}\left(d,\vec{\tau}\right)+u_{t}\left(d,\vec{\tau},\vec{r},\theta\right)
\end{equation}
where $u_{cz}$ is the reference energy in the absence of a twist, and can be expressed using a form such as Eq \ref{eq:xuber}. 
The function $u_{t}\left(d,\vec{\tau},\vec{r},\theta\right)$ accounts for the twist energy and is equal to zero when $\theta$=0.
While the $\theta$ dependence of $u_{t}\left(d,\vec{\tau},\vec{r},\theta\right)$ could be represented as a Fourier series, as it is periodic in $\theta$, it may exhibit cusps and therefore not be a smooth function of $\theta$ \cite{sutton1987overview,sutton1995}. 
The expansion coefficients of such a Fourier series would be a function of $d$, $\vec{\tau}$ and $\vec{r}$.

\section{Creating slab models}\label{sec:slab}

Most electronic structure methods impose periodic boundary conditions on the crystallographic model. 
In this section, we describe how crystallographic models can be constructed to realize crystal separation, glide and twist within super cells that have periodic boundary conditions. 

\subsection{Constructing slab geometries to parameterize cohesive zone models}\label{sec:uber}

\begin{figure}
\centering
\subfloat[]{
   \includegraphics[width=0.7\linewidth]{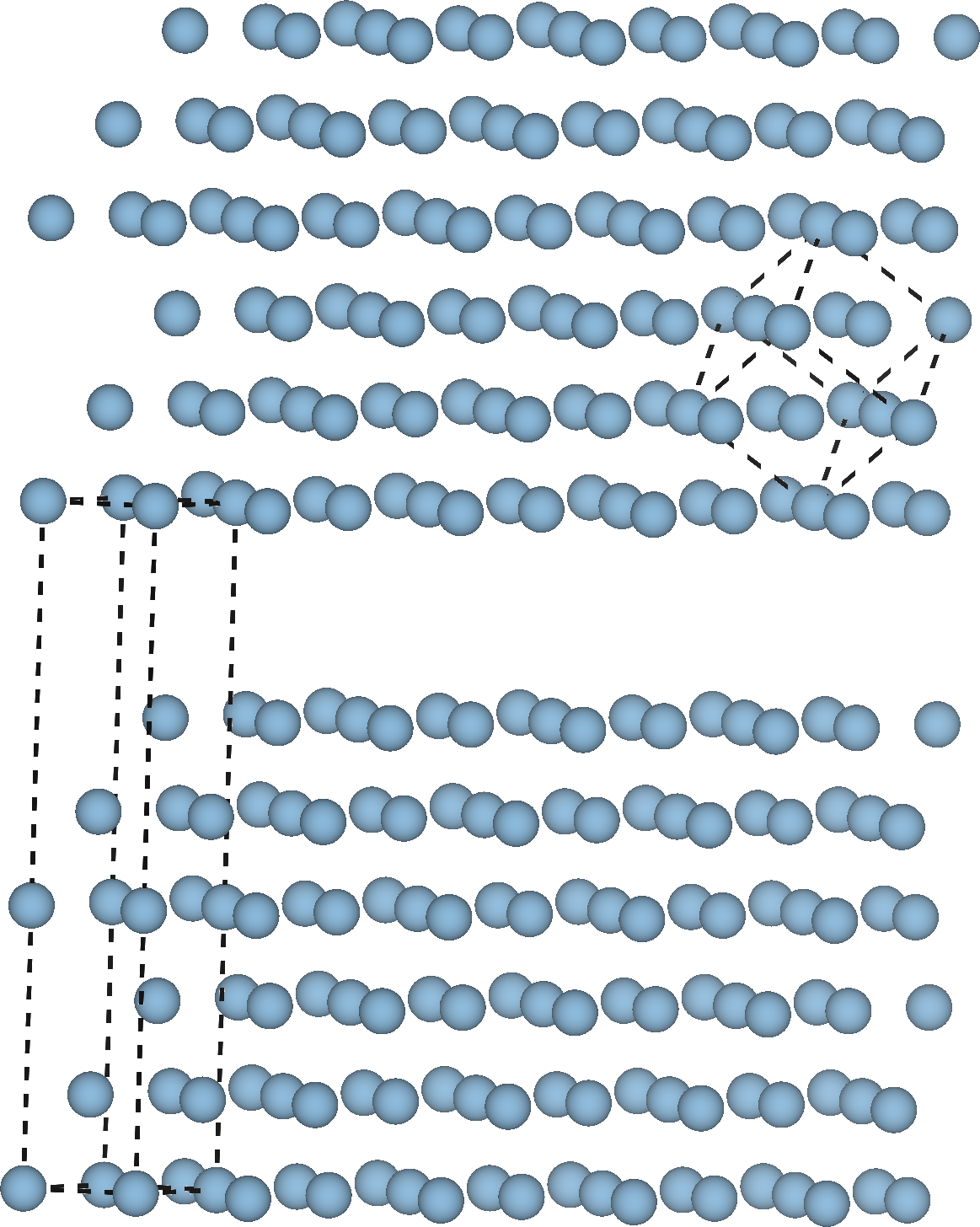}
    \label{fig:slab}
}
\hfill
\subfloat[]{
    \includegraphics[width=\linewidth]{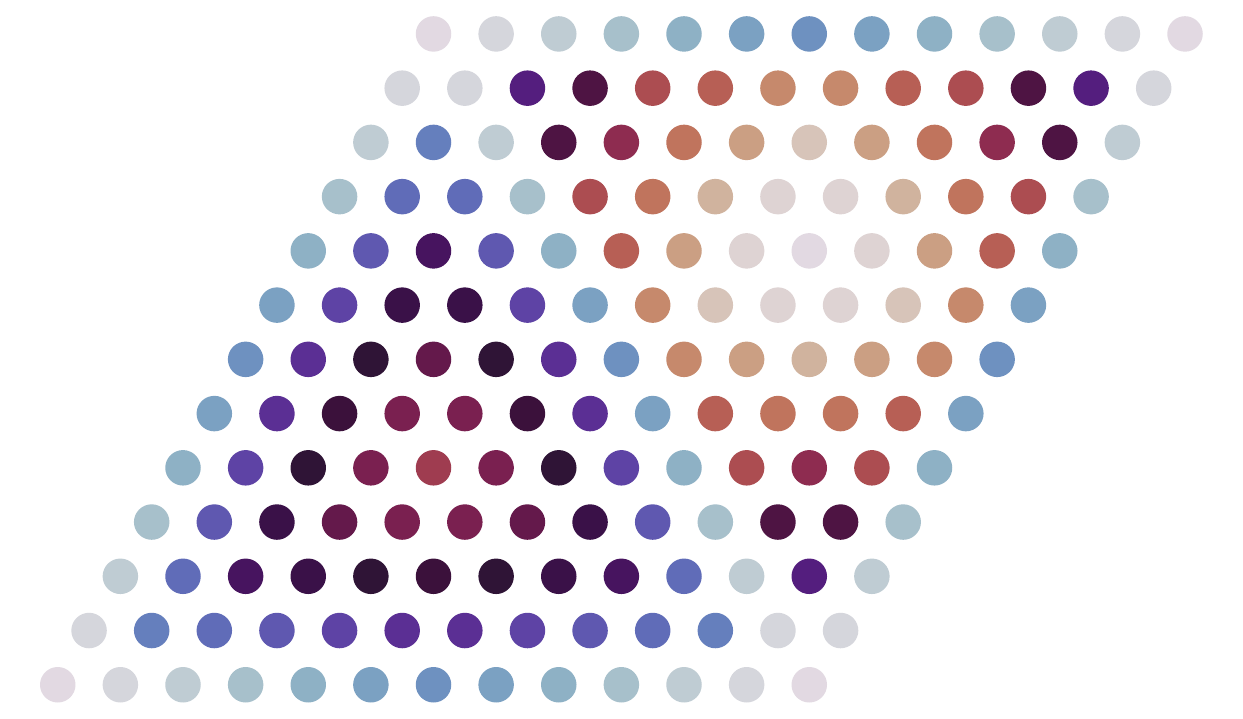}
    \label{fig:shiftsym}
}
\caption{(a) A slab of fcc that has been cleaved along the (1,1,1) plane.
    The unit cell of the slab has had 3\AA{} of vacuum inserted above the exposed plane.
    The conventional unit cell of fcc is also shown for reference. (b) Symmetric equivalence of translation vectors parallel to the (1,1,1) plane of fcc.
    Spots with the same color correspond to translation vectors that generate symmetrically equivalent structures.}
\label{fig:slabmodel} 
\end{figure}

A cohesive zone model such as \Cref{eq:xuber} can be parameterized by fitting to training data as calculated with a first-principles electronic structure method. 
It is possible to accommodate the periodic boundary conditions that these methods impose using a slab geometry as illustrated in \Cref{fig:slab}. 
The unit cell then consists of a slab of crystal with its periodic images separated by layers of vacuum parallel to the plane {\bf P}.
For bulk crystals, the crystal slab must be sufficiently thick to avoid interactions between periodic images of the surfaces adjacent to the vacuum layer. 

To construct a crystallographic model consisting of slabs parallel to a plane {\bf P}, it is necessary to first identify a super cell of the primitive cell vectors, $\vec{a}$, $\vec{b}$ and $\vec{c}$, such that two super cell vectors, $\vec{A}$ and $\vec{B}$ span the plane {\bf P}.
The vectors $\vec{A}$ and $\vec{B}$ can be determined by connecting the origin to the closest non-collinear lattice points that lie on a plane parallel to {\bf P} that also passes through the origin. 
A third vector, $\vec{s}$, can then be chosen as the shortest translation vector of the parent crystal structure that is not in the plane {\bf P}. 
The vector $\vec{s}$ determines the smallest possible thickness of the slab.
The thickness of the slab can be adjusted by multiplying $\vec{s}$ by an integer, $l$. 
It is usually desirable to translate the resulting vector $l\vec{s}$ parallel to {\bf P} (by an integer linear combination of $\vec{A}$ and $\vec{B}$) until its projection onto the plane {\bf P} falls within the unit cell spanned by $\vec{A}$ and $\vec{B}$.
This new vector, $\vec{C}$, is then the third super cell vector of the slab model. 

The next task is to sample different values of slab separations $d$ and relative translations $\vec{\tau}$. 
Due to translational periodicity of the crystal, only translation vectors $\vec{\tau}$ within a two-dimensional unit cell spanned by the $\vec{A}$ and $\vec{B}$ vectors of the super cell need to be considered. 
One approach is to generate a uniform grid within the unit cell of possible translation vectors $\vec{\tau}$, as is illustrated in \Cref{fig:shiftsym} for an fcc crystal in which one half is sheared relative to another half along a (111) plane. 
The two half crystals often have additional symmetries that make a subset of the translations $\vec{\tau}$ equivalent to each other. 
Symmetric equivalence between two different translation vectors $\vec{\tau}_1$ and $\vec{\tau}_2$ can be ascertained by mapping the resultant crystals onto each other with a robust crystal mapping algorithm \cite{mapping_paper}. 
\Cref{fig:shiftsym} shows the orbits of equivalent translation vectors $\vec{\tau}$ for fcc by assigning the same color to all translation vectors that are equivalent by symmetry. 
The choice of super cell may break some symmetries of the underlying crystal, making symmetric equivalence of translation vectors $\vec{\tau}$ dependent on the particular super cell. 
For each symmetrically distinct translation vector $\vec{\tau}$ it is possible to generate a grid of separations $d$ over increments of $\Delta d$. 
This can be realized by adding $\Delta d \mathbf{\hat{n}}$ to $\vec{C}$ while keeping the Cartesian coordinates of the atoms within the unit cell unchanged. The vector $\mathbf{\hat{n}}$ is a unit vector normal to the $\vec{A}$-$\vec{B}$ plane. 

The parameterization of a generalized cohesive zone model can occur in two steps.
The first step is to calculate the energy of separation over a discrete set of separations $d$ for each symmetrically distinct translation vector $\vec{\tau}$. 
The resultant energy versus separation $d$ relation can then be fit to an xUBER relation, Eq \ref{eq:xuber}. 
The parameterization of an xUBER relation over all symmetrically distinct translation vectors $\vec{\tau}$ will generate numerical values for the adjustable parameters, $d_0$, $\gamma$, $\lambda$ and $\alpha_{n}$, of the xUBER, \Cref{eq:xuber}, over a uniform grid of translation vectors $\vec{\tau}$.
The dependence of the adjustable parameters on $\vec{\tau}$ can then be represented with a Fourier series such as Eq. \ref{eq:fourier}, thereby allowing for the accurate interpolation at any translation vector $\vec{\tau}.$

\subsection{Crystallographic models for twisted bicrystals}\label{sec:bicrystals}

\begin{figure}
\centering
\subfloat[]{
    \includegraphics[width=\linewidth]{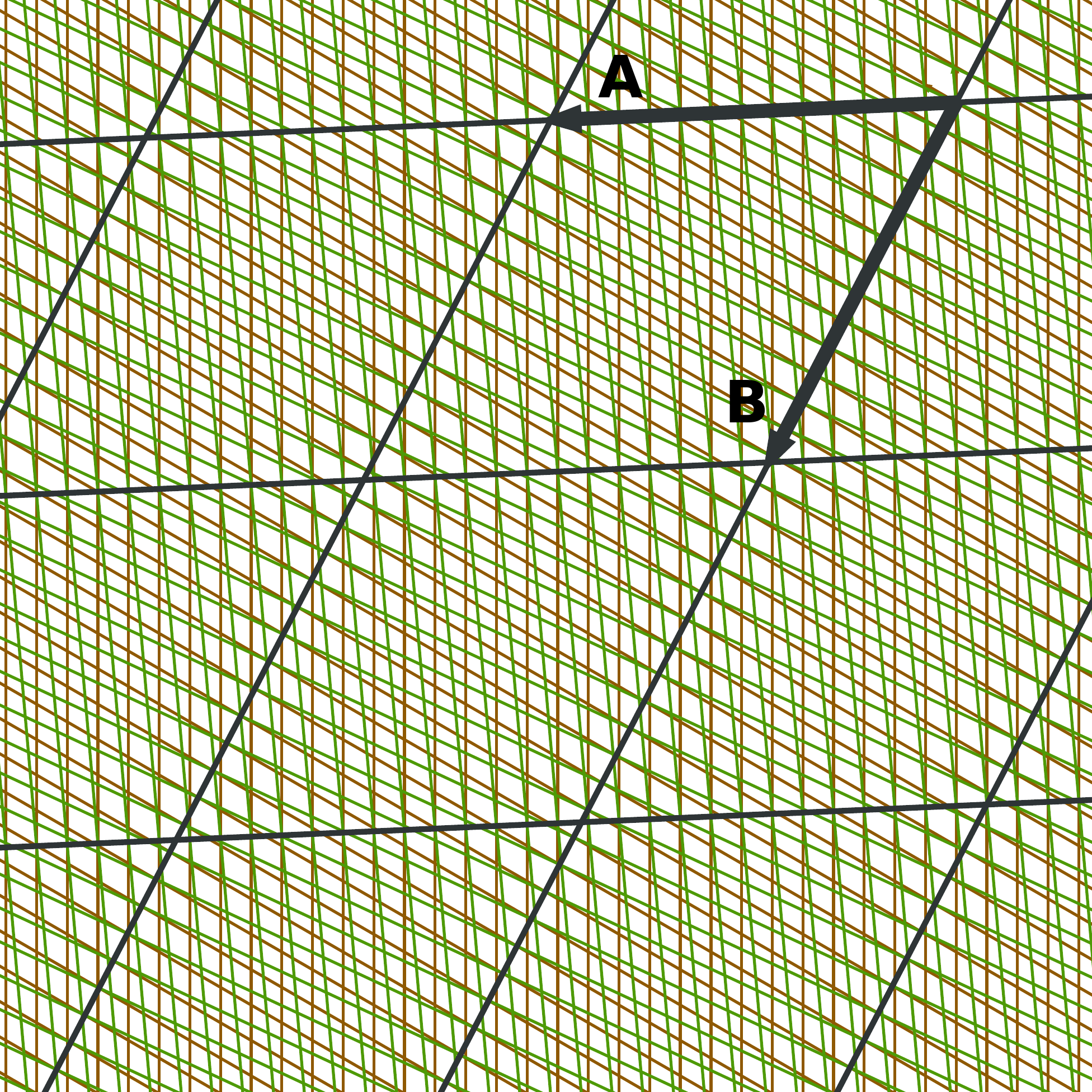}
    \label{fig:moire5}
}

\centering
\subfloat[]{
    \includegraphics[width=0.9\linewidth]{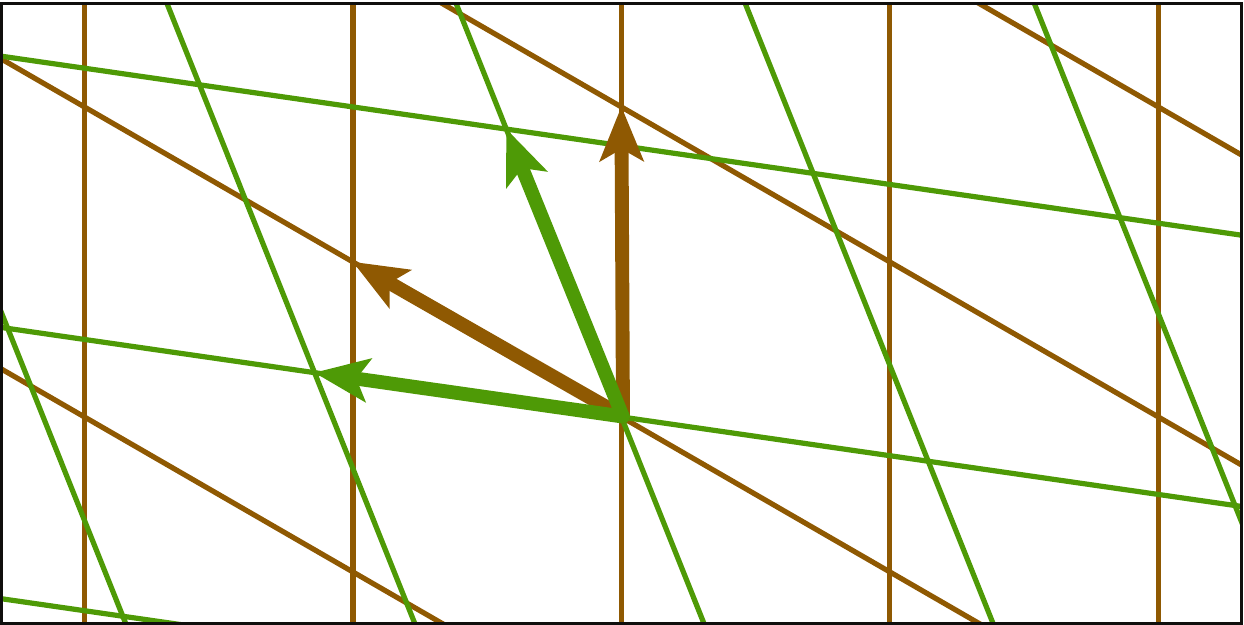}
    \label{fig:rot21.79real}
}
\hfill
\subfloat[]{
    \includegraphics[width=0.9\linewidth]{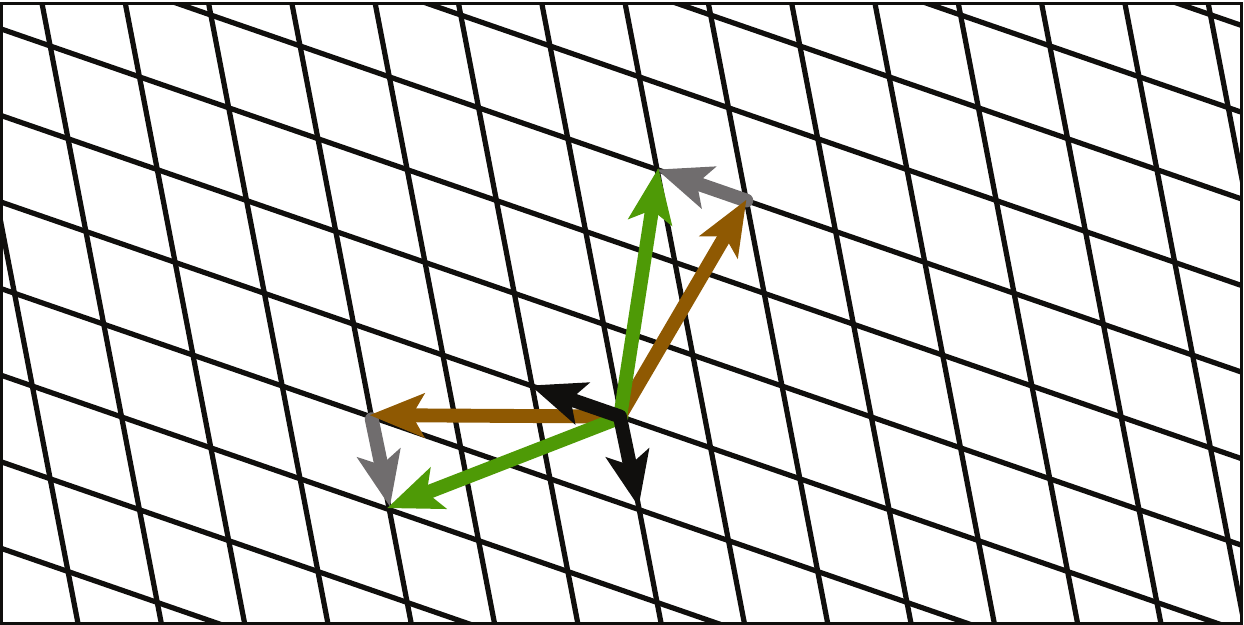}
    \label{fig:rot21.79recip}
}
\caption{(a) Superposition of two triangular lattices in which one (green) has been rotated 5 degrees relative to the other one (brown). The resulting interference pattern forms the Moir\'e lattice. (b) The reciprocal lattices of the two triangular lattices of (a).
(c) The reciprocal lattice of the Moir\'e lattice (black vectors) can be constructed by taking the difference between the reciprocal lattice vectors of the rotated lattices (shown as gray vectors).}
\label{fig:moireconstruct} 
\end{figure}

In addition to separation and glide, it is also possible to twist the two halves of a bicrystal by rotating one half relative to the other around a rotation axis, $\vec{r}$, that is perpendicular to {\bf P}.
It is increasingly recognized that interesting electronic properties can emerge when a pair of two-dimensional materials are rotated relative to each other in this manner \cite{cao2018correlated,cao2018unconventional,burch2018magnetism,carr2020electronic}. 
In bulk materials, special grain boundaries, referred to as twist boundaries, can be generated by twisting the top half of a crystal around an axis that is perpendicular to the grain boundary \cite{sutton1995}.
A challenge for electronic structure calculations is to identify a super cell that is able to accommodate the twisted half crystals. 

A Moir\'e pattern emerges when a periodic two-dimensional lattice is rotated relative to a second periodic lattice. 
Figure \ref{fig:moire5}, for example, shows the emergence of a Moir\'e pattern after a pair of two-dimensional triangular lattices (brown and green) have been rotated relative to each other by 5$^{\circ}$. 
The Moir\'e pattern is itself periodic, but has lattice vectors that are usually much larger than those of the two-dimensional lattices that have been rotated. 
Furthermore, the lattice of the Moir\'e pattern is rarely commensurate with the lattices of the twisted half crystals.
This is evident in Figure \ref{fig:moire5}, where the lattice points of the Moir\'e pattern, shown as the intersections of the grey lines, do not exactly overlap with sites of the twisted triangular lattices shown in brown and green. 
Only a subset of particular twist angles $\theta$ produce Moir\'e patterns that are commensurate with the twisted two-dimensional lattices \cite{shallcross2008quantum,shallcross2010electronic,mele2010commensuration,carr2020electronic}.
The Moir\'e lattice can, nevertheless, serve as a guide to identify a super cell that can accommodate the twisted half crystals for first-principles electronic structure calculations. 
But since the Moir\'e lattices for most twist angles $\theta$ only approximately coincide with sites of the twisted lattices, both half crystals must usually be deformed slightly such that they can be accommodated within a common super cell. 

The lattice of the Moir\'e pattern can be determined by working in reciprocal space. 
Consider a three dimensional unit cell with lattice vectors $\vec{A}$, $\vec{B}$ and $\vec{C}$. 
Assume that the vectors $\vec{A}$ and $\vec{B}$ form the two-dimensional lattice parallel to the plane {\bf P} (e.g. a two-dimensional triangular lattice) and that the third vector $\vec{C}$ is perpendicular to {\bf P}.
The rotation axis $\vec{r}$ is therefore parallel to $\vec{C}$.
It is convenient to work with a $3\times3$ matrix $\mat{L} = [\vec{A}, \vec{B}, \vec{C}]$ where each lattice vector appears as a column of $\mat{L}$.  
The reciprocal lattice vectors of the lattice $\mat{L}$ are then the column vectors of the matrix $\mat{K}$ defined by 
\begin{equation}
    \mat{K}=2\pi \left( \mat{L}^{-1} \right)^\intercal 
    \label{eq:reciprocal}
\end{equation}
The application of a rotation $\theta$ to a lattice $\mat{L}$ around a rotation axis that is parallel to $\vec{C}$ produces a lattice  $\mat{L}_{\theta}$.
The lattice vectors of the two-dimensional Moir\'e pattern, $\mat{M}$, will have a reciprocal lattice represented by the matrix $\mat{K}_{M}$ in which the upper left $2\times2$ block is equal to the corresponding difference between the reciprocal lattices of $\mat{L}$ and $\mat{L}_{\theta}$ (i.e. $\left[\mat{K}_{M}\right]_{i,j} = \left[\mat{K}\right]_{i,j} - \left[\mat{K}_{\theta}\right]_{i,j}$ for $i,j = 1,2$) and the third axis, which is unaffected by the rotation is the same as that of $\mat{K}$ and $\mat{K}_{\theta}$ (i.e. $\left[\mat{K}_{M}\right]_{3,3}=\left[\mat{K}\right]_{3,3}=\left[\mat{K}_{\theta}\right]_{3,3}$).   
The Moir\'e lattice is then 
\begin{equation}
    \mat{M}=2\pi\left( \mat{K}_{M}^{-1}\right)^{\intercal}
    \label{eq:moire}
\end{equation}
This is illustrated in \Cref{fig:rot21.79real} and \Cref{fig:rot21.79recip} for a pair of triangular lattices. The green triangular lattice of \Cref{fig:rot21.79real} has been rotated by an angle $\theta$ relative to the brown triangular lattice. The reciprocal lattice vectors of the green and brown triangular lattices are shown in \Cref{fig:rot21.79recip}. The two grey vectors in \Cref{fig:rot21.79recip} are the differences of the reciprocal lattice vectors of the rotated green lattice and those of the fixed brown lattice. 
The grey vectors, when translated to the origin of reciprocal space, span the unit cell of the reciprocal lattice of the Moir\'e pattern. 
This is illustrated by the thick black arrows in \Cref{fig:rot21.79recip} with the grid representing the reciprocal lattice points of the Moir\'e lattice. 
(For large rotation angles $\theta$, it is possible that one of the reciprocal lattice vectors of $\mat{K}_{M}$ falls outside of the Wigner-Seitz cell of either $\mat{K}$ or $\mat{K}_{\theta}$. 
In these situations, the offending reciprocal lattice vector of $\mat{K}_{M}$ must be translated back into the Wigner-Seitz cell of either $\mat{K}$ or $\mat{K}_{\theta}$.) 

The example illustrated by \Cref{fig:rot21.79real} and \Cref{fig:rot21.79recip} is for a special angle $\theta$ for which the Moir\'e lattice is commensurate with the two rotated lattices. 
For these special angles, the reciprocal lattice vectors $\mat{K}$ and $\mat{K}_{\theta}$ (brown and green vectors in \Cref{fig:rot21.79recip}) coincide with sites of the reciprocal lattice of the Moir\'e lattice $\mat{K}_{M}$.
Only a subset of twist angles produce Moir\'e lattices that are commensurate with the twisted lattices\cite{carr2020electronic}.

In general, the Moir\'e lattice points do not exactly coincide with sites of either the $\mat{L}$ or $\mat{L}_{\theta}$ lattices, as illustrated by the example of \Cref{fig:moire5} for a pair of triangular lattices that have been rotated by 5$^{o}$.
Nevertheless, the Moir\'e lattice can guide the search for a super cell that can simultaneously accommodate the twisted pair of two-dimensional lattices. 
The first task is to identify the super cells of $\mat{L}$ and $\mat{L}_{\theta}$ that are close to that of the Moir\'e lattice. 
A super cell of a lattice can be generated as an integer linear combination of the lattice vectors $\vec{A}$, $\vec{B}$, $\vec{C}$ according to
\begin{equation}
    \mat{S}=\mat{L}\mat{T}
    \label{eq:transfmat}
\end{equation}
where $\mat{T}$ is a $3\times3$ integer matrix and where the columns of $\mat{S}$ contain the super cell lattice vectors. 
The sought after integer matrix $\mat{T}$ is one that generates a super cell $\mat{S}$ that is closest to that of the Moir\'e pattern $\mat{M}_{\theta}$.
This can be obtained by rounding each element of the matrix $\mat{L}^{-1}\mat{M_{\theta}}$ to the nearest integer. 
A similar matrix $\mat{T}_{\theta}$ must be determined by rounding the elements of $\mat{L}_{\theta}^{-1}\mat{M_{\theta}}$ to the nearest integer. 
The resulting super cells, $\mat{S}=\mat{L}\mat{T}$ and $\mat{S}_{\theta}=\mat{L}_{\theta}\mat{T}_{\theta}$, will usually not coincide exactly. 
However, they can both be strained and twisted slightly to a common super cell $\mat{\bar{S}}$ defined as the average of $\mat{S}$ and $\mat{S}_{\theta}$ according to
\begin{equation}
    \mat{\bar{S}}=\frac{1}{2}\left(\mat{S}+\mat{S}_{\theta}\right)
\end{equation}
This super cell can be used to accommodate the twisted bicrystal.

The amount of strain and twist needed to fit both bicrystals in $\mat{\bar{S}}$ can be calculated as follows.
The dimensions of the bottom half of the bicrystal with superlattice $\mat{S}$ will need to be deformed according to 
\begin{equation}
    \mat{F}\mat{S}=\mat{\bar{S}}
\end{equation}
where $\mat{F}$, the deformation gradient, is a $3\times 3$ matrix. 
The deformation gradient $\mat{F}$ can be factored into a product of a symmetric stretch matrix $\mat{U}$ and a rotation matrix $\mat{R}$ according to $\mat{F}=\mat{R}\mat{U}$. The stretch matrix $\mat{U}$ describes the deformation of the crystal, while the rotation matrix $\mat{R}$ in this situation corresponds to a rotation around $\vec{C}$.
A similar deformation gradient exists for the top half of the bicrystal with $\mat{F}_{\theta}\mat{S}_{\theta}=\mat{\bar{S}}$ and $\mat{F}_{\theta}=\mat{R}_{\theta}\mat{U}_{\theta}$. 
If the rotation angles of $\mat{R}$ and $\mat{R}_{\theta}$ are $\phi_{1}$ and $\phi_2$, respectively, then the two bicrystals need to undergo an additional relative twist of $\Delta \theta = \phi_2-\phi_1$ to fit into $\mat{\bar{S}}$.
The actual rotation angle when describing the twisted bicrystal with a periodic super cell $\mat{\bar{S}}$ is then $\theta +\Delta \theta$.
Hence, it is generally not possible to realize the target twist angle of $\theta$ when using a periodic super cell to accommodate the twisted halves. 

Information about the strains is embedded in the stretch matrices $\mat{U}$ for the bottom half and $\mat{U}_{\theta}$ for the top half.
We use the Biot strain defined as $\mat{E}=\mat{U}-\mat{I}$ where $\mat{I}$ is the identity matrix \cite{thomas2017exploration}. 
The strain is restricted to the two-dimensional space parallel to the twist plane.
We assume that this plane is parallel to the $\hat{x}$-$\hat{y}$ plane of the Cartesian coordinate system used to represent the lattice vectors. 
Convenient metrics of the degree with which the two bicrystals are strained is the square root of the sum of the squares of the eigenvalues of the strain matrices $\mat{E}$ and $\mat{E}_{\theta}$ (i.e. $\sqrt{\lambda_1^{2}+\lambda_{2}^{2}}$ where $\lambda_1$ and $\lambda_{2}$ are the non zero eigenvalues of the strain matrices).


\begin{figure}
    \centering
    \includegraphics[width=\linewidth]{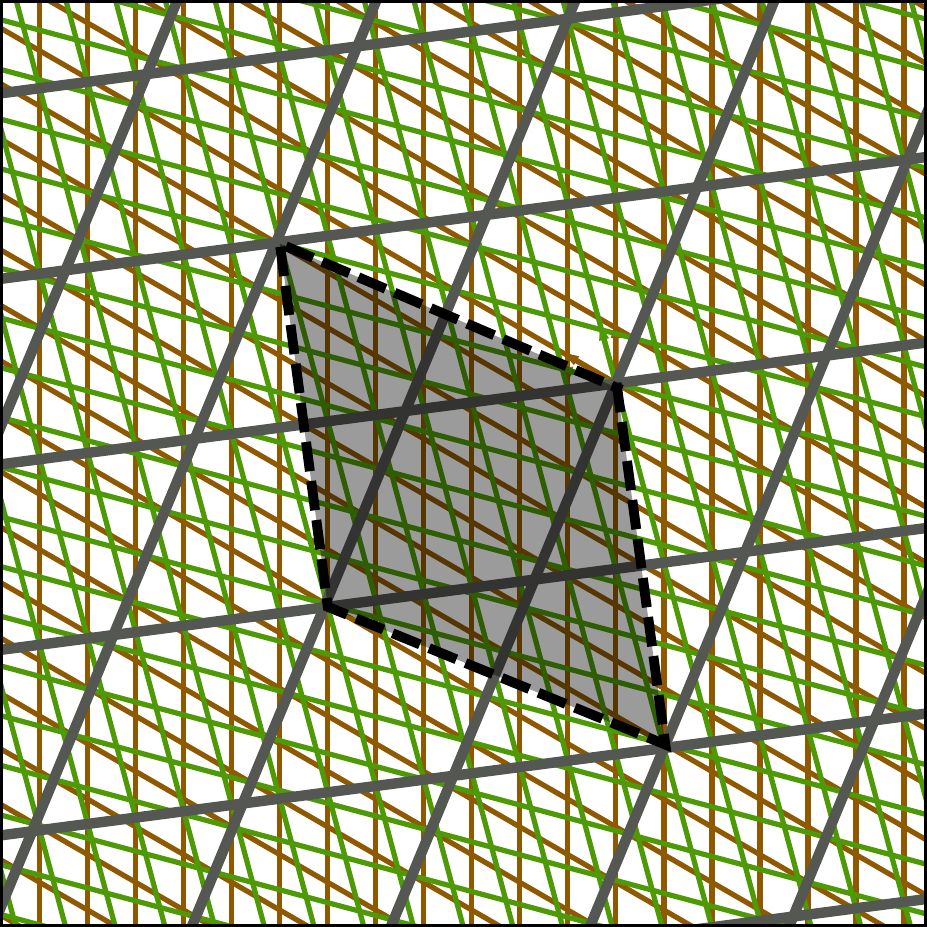}
    \caption{The Moir\'e lattice (gray) that emerges when a pair of triangular lattices have been rotated by 15 degrees, and one of its possible super cells (black).
    By using a super cell of the Moir\'e lattice, it is possible to generate a crystallographic model to accommodate a twisted bicrystal that requires less deformation.}
    \label{fig:supermoire}
\end{figure}

\begin{figure}
    \centering
    \includegraphics[width=\linewidth]{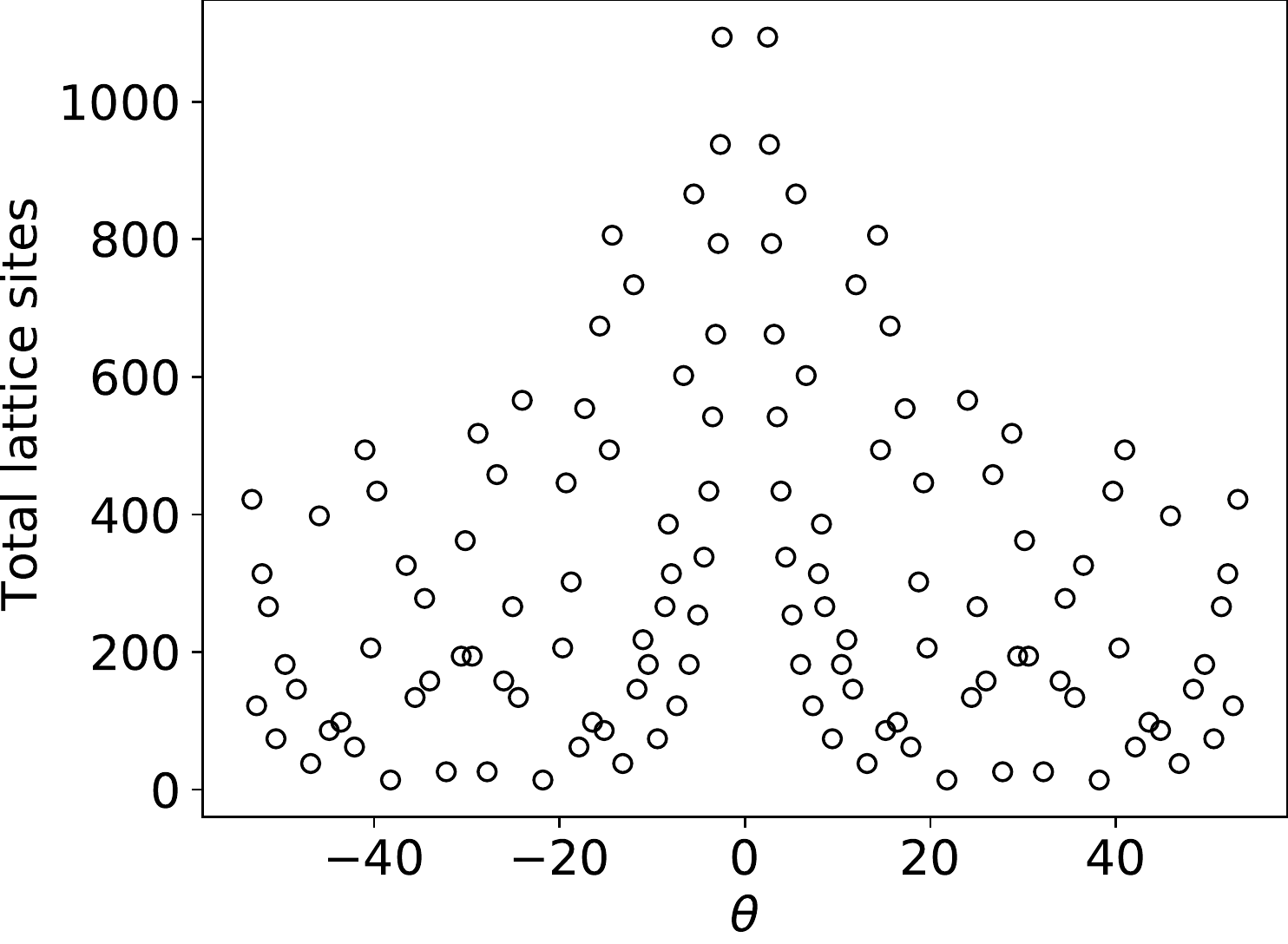}
    \caption{Number of triangular lattice sites within the super cells of bilayers with a commensurate twist angle.}
    \label{fig:totallatticesites1000}
\end{figure}

A refinement of the above approach can be used to lower the error in the target angle $\Delta \theta$ and the incurred strains. 
However, the improvement comes at the cost of requiring larger super cells. 
Instead of identifying super cells $\mat{S}$ and $\mat{S}_{\theta}$ of the lattices $\mat{L}$ and $\mat{L}_{\theta}$ that are as close as possible to the Moir\'e lattice $\mat{M}$, the super cells $\mat{S}$ and $\mat{S}_{\theta}$ can be matched to {\it super cells} of $\mat{M}$. 
This is illustrated in \Cref{fig:supermoire} for a pair of triangular lattices that are being rotated by a target angle of $\theta = 15^{\circ}$. 
When matching super cells $\mat{S}$ and $\mat{S}_{\theta}$ to the Moir\'e lattice $\mat{M}$ itself, as described above, the actual twist angle is $\theta = 12.117^{\circ}$.
In contrast, when matching the super cells $\mat{S}$ and $\mat{S}_{\theta}$ to a $\sqrt{3}a\times\sqrt{3}a$ super cell of the Moir\'e lattice $\mat{M}$ the actual rotation angle becomes 14.911$^{\circ}$, which is much closer to the target angle of 15$^{\circ}$. 
However, the common super cell that accommodates the twisted bicrystal is three time larger. 
There are special twist angles for which super cells of their Moir\'e lattice can be found that are commensurate with the under lying lattices of the bicrystals. For these special twist angles $\Delta \theta$ and the strain order parameters will be zero. 
\Cref{fig:totallatticesites1000} plots the number of triangular lattice unit cells that are needed in the super cells of a subset of commensurate twist angles. 
It is clear that very large super cells are necessary for most commensurate twist angles. 

\section{\mush{}}\label{sec:mush}

\begin{figure}
    \centering
    \includegraphics[width=0.9\linewidth]{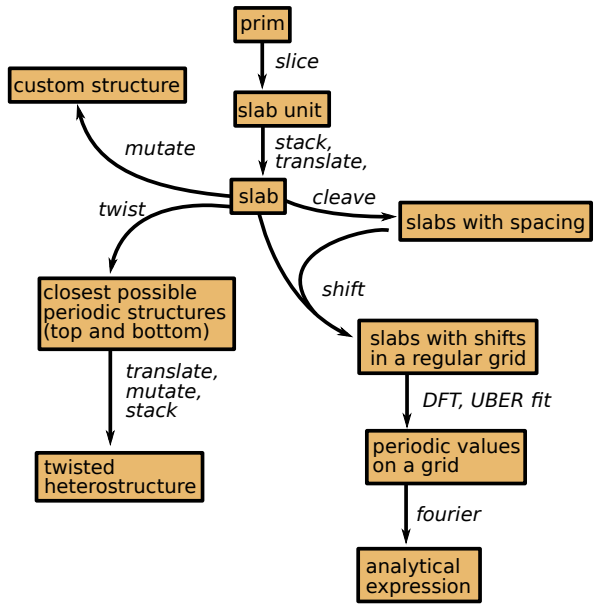}
    \caption{Flowchart of the possible \mush{} commands}
    \label{fig:flowchart}
\end{figure}

The \mush{} software package constructs crystallographic models of extended two-dimensional defects that separate a pair of bicrystals as described in the previous section. 
This includes crystallographic models to parameterize generalized cohesive zone models such as Eq. \ref{eq:xuber} and crystallographic models of twisted bicrystals. 
\Cref{fig:flowchart} illustrates a schematic flowchart of \mush{}.

The first step of the \mush{} workflow is to construct the slab building blocks. 
The user provides the primitive cell of a crystal ({\it prim} in \Cref{fig:flowchart}) and the Miller indices of the crystallographic plane {\bf P}.
In the {\it slice} step, \mush{} constructs the thinnest super cell of {\it prim} with lattice vectors $\vec{A}$ and $\vec{B}$ parallel to the plane {\bf P} and the shortest vector $\vec{C}$ that is not in the plane {\bf P}.
The resulting super cell contains the {\it slab unit} (\Cref{fig:flowchart}) that constitutes the fundamental building block of subsequent crystallographic models. 
The slab unit must next be made thicker when modeling three dimensional materials or stacked when modeling the twist of two-dimensional materials. 
The slab thickness is increased based on user input. 
This occurs with the {\it stack} step.
Depending on the basis of the crystal, an additional {\it translate} may be required, where the basis atoms in the slab unit are rigidly shifted, changing the position through which the plane $\mat{P}$ penetrates the crystal.
For example, when exploring the glide of CoO$_2$ sheets in a layered battery material such as $\mathrm{LiCoO_2}$, $\mat{P}$ should extend between the oxide layers, not through them.

At this point, the \mush{} workflow diverges into two tracks. 
The first (right arrows), generates crystallographic models of crystal cleavage and glide with respect to a plane {\bf P}.
A list of separations $d$ is specified ({\it cleave}) and for each of these separations a second grid of glide vectors $\vec{\tau}$ may be enumerated ({\it shift}). 
Symmetrically equivalent translation grid points are tagged as such. 
Directories with input files for the VASP\cite{kresse1993ab,kresse1994ab,kresse1996efficiency,kresse1996efficient} first-principles electronic structure software package are then generated. 
Upon completion of static electronic structure calculations, a list of first-principles energies are available to paramaterize a generalized cohesive zone model.

An alternative to the imposed regular grid of translation vectors is to use the {\it mutate} step.
With this approach, a single custom structure that has been shifted by an arbitrary vector $\vec{\tau}$ and separated by a custom value $d$ from its periodic image is created.

A second track in the \mush{} workflow (left arrows) generates crystallographic models of twisted bicrystals. 
Here the user provides a target twist angle $\theta$ and a maximum number of two-dimensional unit cells for the super cell that will accommodate the twisted bicrystals. 
\mush{} next determines the Moir\'e lattice and then identifies the super cells of two bicrystals that best match a super cell of the Moir\'e lattice.
A crystallographic model is output along with the actual twist angle $\theta+\Delta \theta$ and values of the strain order parameters $\eta_1$, $\eta_2$ and $\eta_3$. 
When twisting a pair of bicrystals within a unit cell constrained by periodic boundary conditions (including the $\vec{C}$ axis) two interfaces are necessarily introduced. 
One is the twist interface of interest, while the other is usually separated by a large slab of vacuum. 
For two-dimensional materials, the vacuum is not necessarily a drawback. 
For twist grain boundaries, however, the thicknesses of the twisted slabs should be sufficiently large such that the free surfaces in contact with vacuum do not affect the energy and electronic structure of the twist boundary. 

\section{Examples}\label{sec:examples}

\begin{figure}
\centering
\subfloat[]{
    \includegraphics[width=0.8\columnwidth]{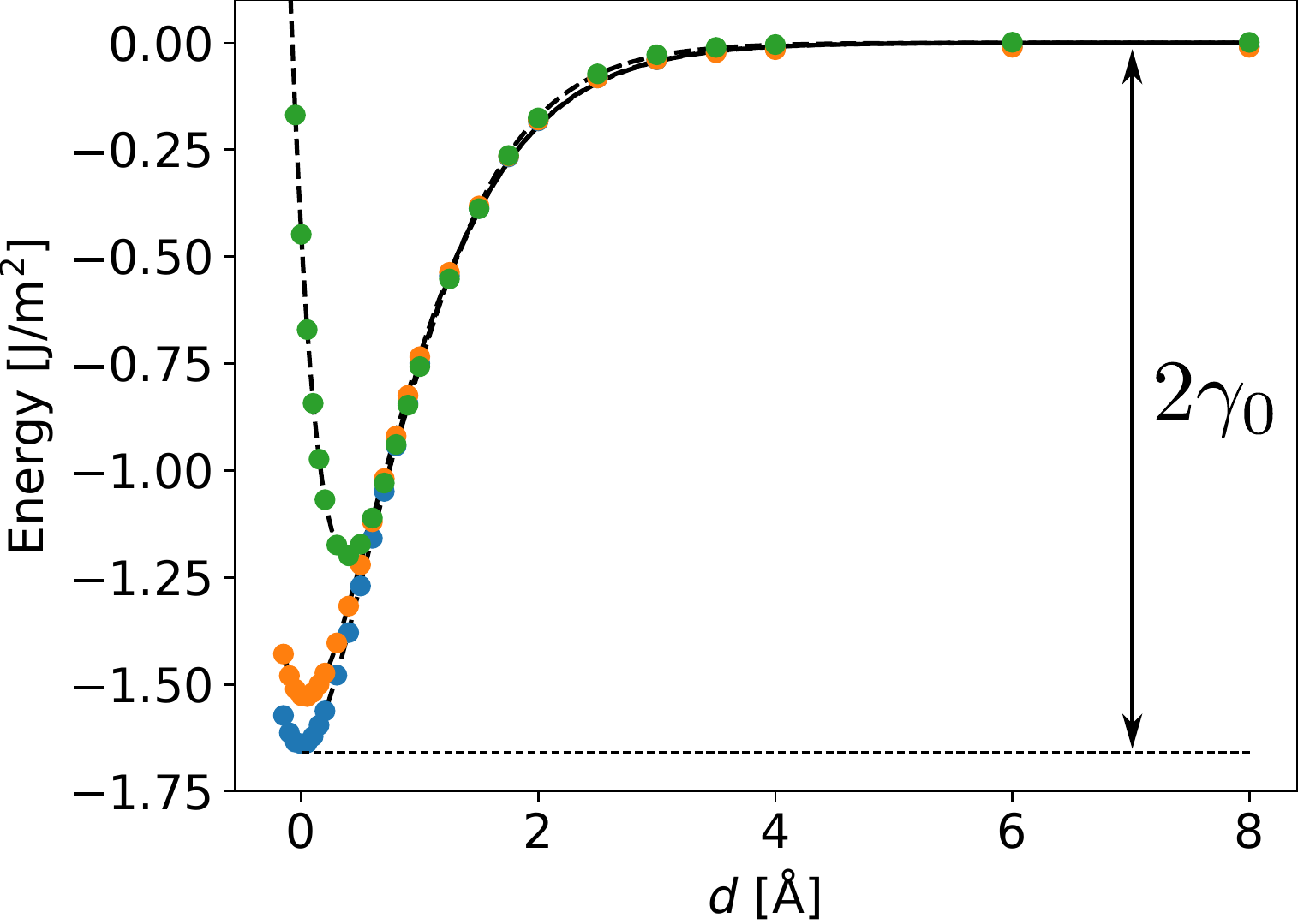}
    \label{fig:uberstack}
}
\hfill
\subfloat[]{
    \includegraphics[width=0.9\linewidth]{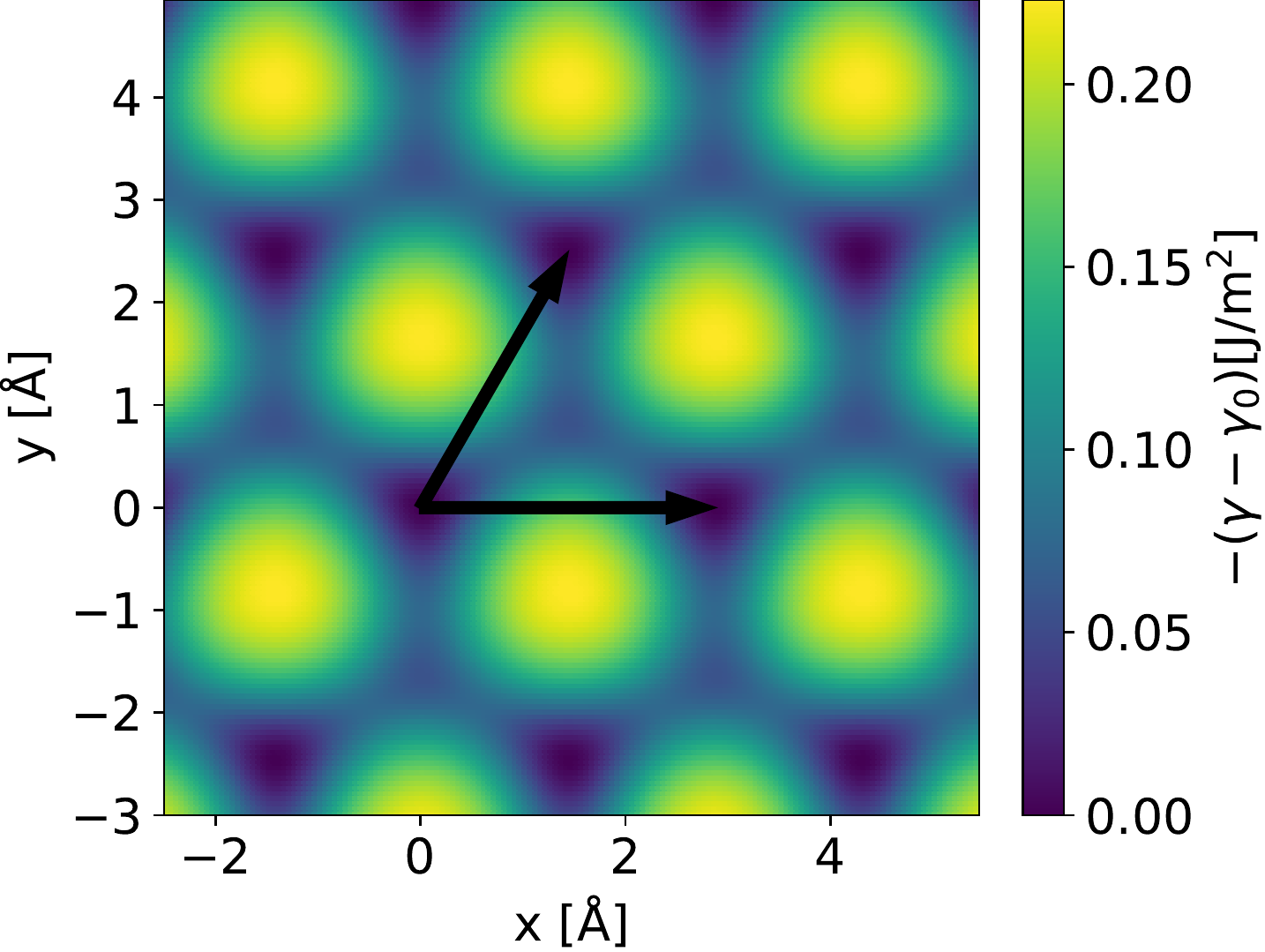}
    \label{fig:al.gammasurf}
}
\hfill
\subfloat[]{
    \includegraphics[width=0.9\linewidth]{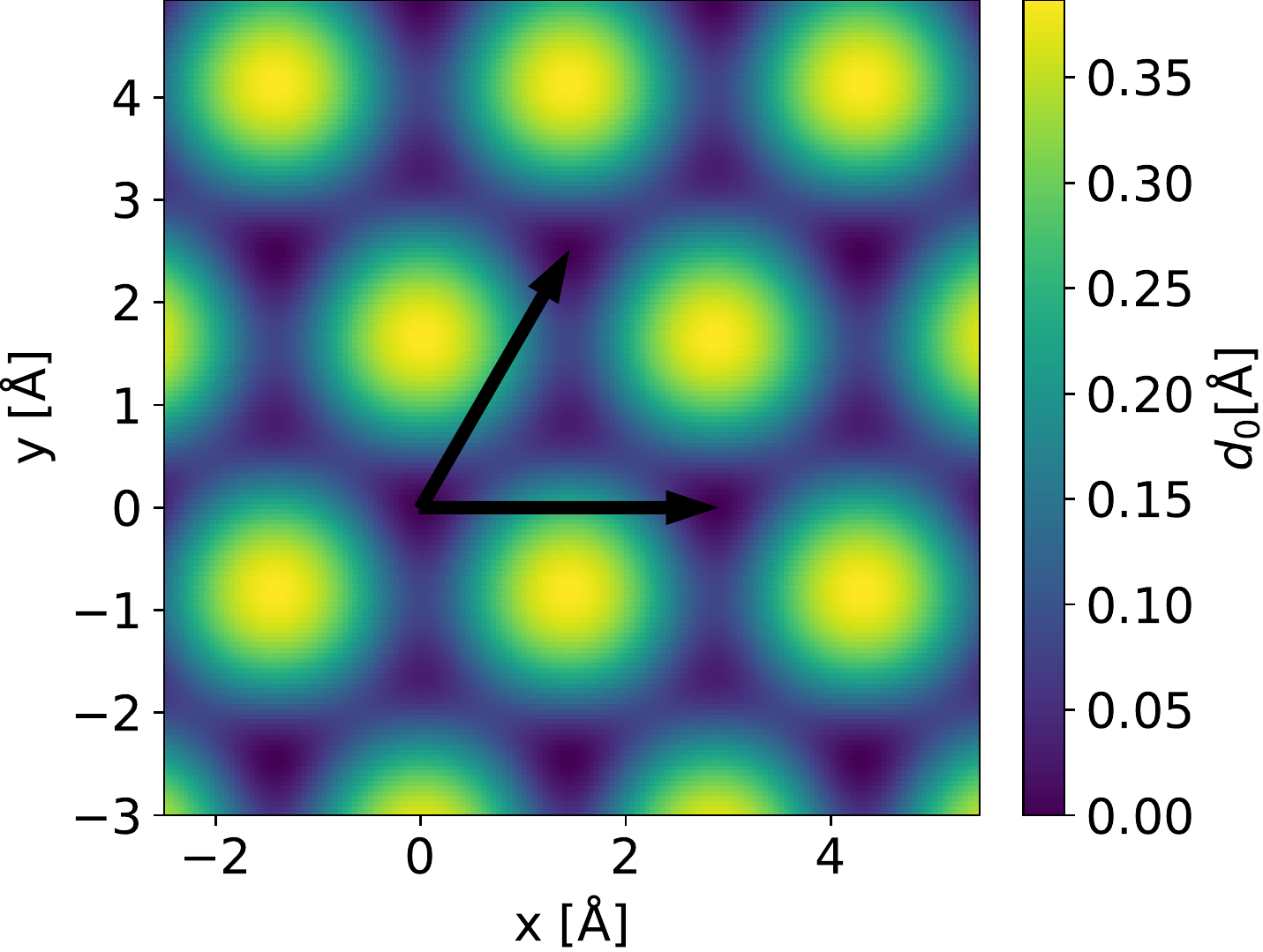}
    \label{fig:al.d0surf}
}

\caption{(a) Energy versus separation of fcc Al as the crystal is separated between a pair of adjacent (111) planes as calculated using the super cell shown in \Cref{fig:slab}. The dashed lines represent UBER fits through the first-principles (DFT) data points. Separation curve for perfect fcc, a stacking fault, and a translation that places a pair of adjacent (111) planes directly on top of each other are shown in blue, orange, and green respectively. (b) Energy of the same Al super cell as a function of glide parallel to the (111) plane evaluated at the $d_0$ separation for each glide vector. (c) The value of $d_0$ as a function of glide (relative to the equilibrium separation of fcc Al).}
\label{fig:al_surfaces} 
\end{figure}

\subsection{Cohesive zone models of simple metals}

As an illustration of a generalized cohesive zone model, we consider cleavage and glide with respect to a (111) plane of fcc Al. 
\Cref{fig:uberstack} shows the energy of an Al crystal as it is cleaved along a pair of neighboring (111) planes for three different translational shifts $\vec{\tau}$. 
The points were calculated with density functional theory (DFT) within the generalized gradient approximation (GGA-PBE) using the VASP plane wave code \cite{kresse1993ab,kresse1994ab,kresse1996efficiency,kresse1996efficient,blochl1994projector,kresse1999ultrasoft}. 
The projector augmented wave (PAW) method \cite{blochl1994projector} was used with a plane-wave energy cutoff of 520 eV. 
K-points were generated using a fully automatic mesh with a length parameter $R_k=50$ \AA{}$^{-1}$. The UBER form of Eq. \ref{eq:xuber} was fit through the DFT points and is shown as dashed lines in \Cref{fig:uberstack}.
As is clear from \Cref{fig:uberstack}, the UBER curve is able to fit the DFT data very well. 
The blue points in \Cref{fig:uberstack} reside on the energy versus separation curve for perfect fcc corresponding to $\vec{\tau}=0$. 
The difference in energy at $d_0$, where the curve has a minimum, and the energy for large separations corresponds to the surface energy $2\gamma$. 
The energy versus separation curve when separating pairs of (111) planes that form a stacking fault corresponding to $\vec{\tau} = 1/3\vec{A}+1/3\vec{B}$ is very similar, as is evident by the orange points in \Cref{fig:uberstack}, although the minimum is not as deep and the equilibrium spacing $d_0$ is slightly shifted to a larger distance. 
The energies of separation for a translation that puts atoms of adjacent (111) planes directly on top of each other (green points in \Cref{fig:uberstack}) is also very well described with an UBER curve. 
The minimum is at a much higher energy compared to other translations. 

By performing similar DFT calculations over a uniform grid of symmetrically distinct translation vectors, it becomes possible to express the $\vec{\tau}$ dependence of the adjustable parameters of the UBER curve as a Fourier series. 
\Cref{fig:al.gammasurf}, for example shows the dependence of $2\gamma$ on the translation $\vec{\tau}$.
It exhibits the periodicity of the (111) plane of fcc Al, with the global minima corresponding to perfect fcc and the other minima corresponding to stacking faults. 
\Cref{fig:al.d0surf} shows the dependence of the equilibrium separation, $d_0$, between two half crystals of fcc Al as a function of $\vec{\tau}$.
This plot looks very similar to the $2\gamma$ surface of \Cref{fig:al.gammasurf}, with the minimum in $d_0$ coinciding with the fcc stacking and the maximum in $d_0$ coinciding with a stacking for which a pair of adjacent (111) planes are directly on top of each other. 

\begin{figure}
\centering
\subfloat[]{
    \includegraphics[width=0.7\linewidth]{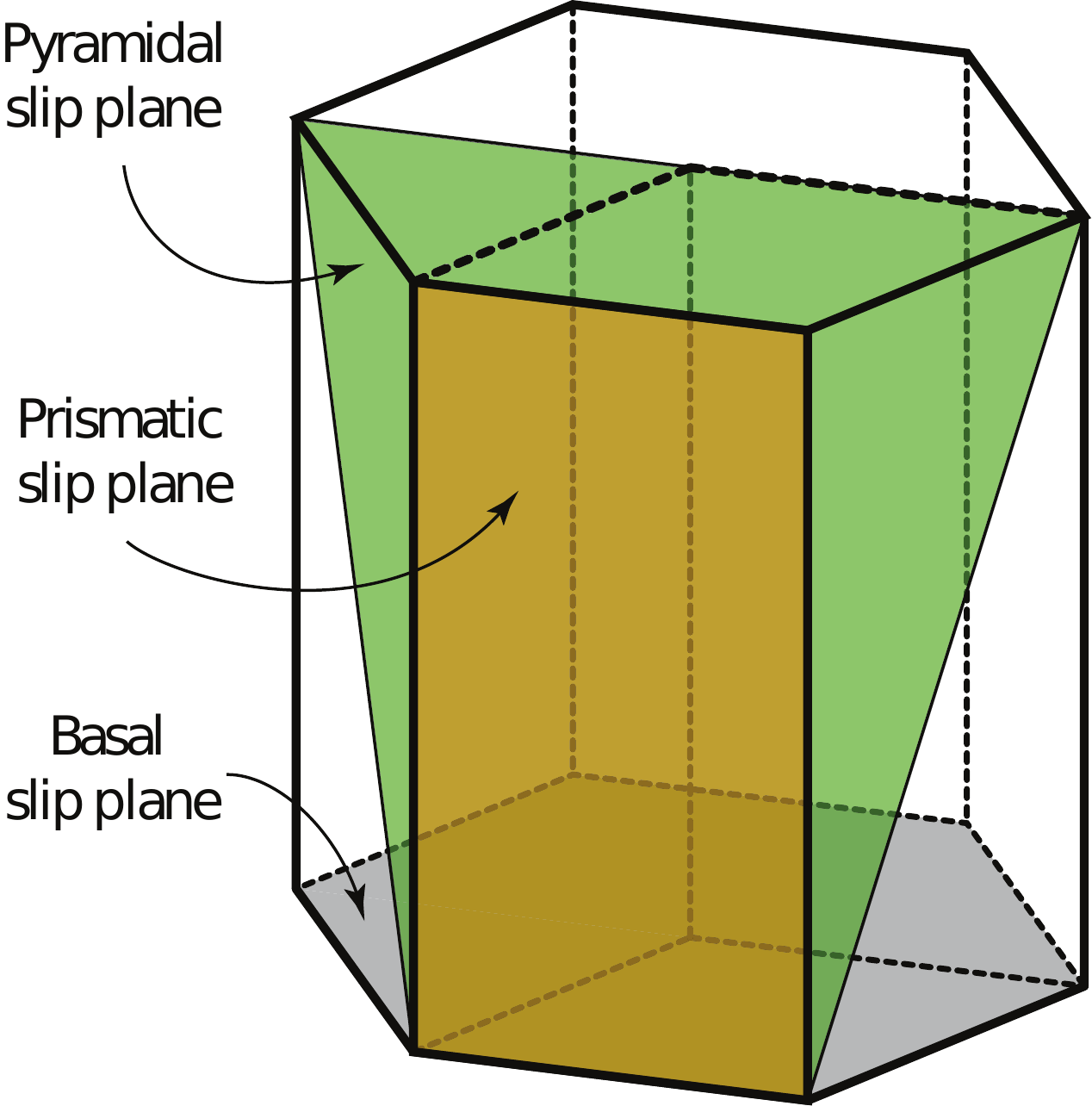}
    \label{fig:hcp_slip_planes}
}
\hfill
\subfloat[]{
    \includegraphics[width=0.9\linewidth]{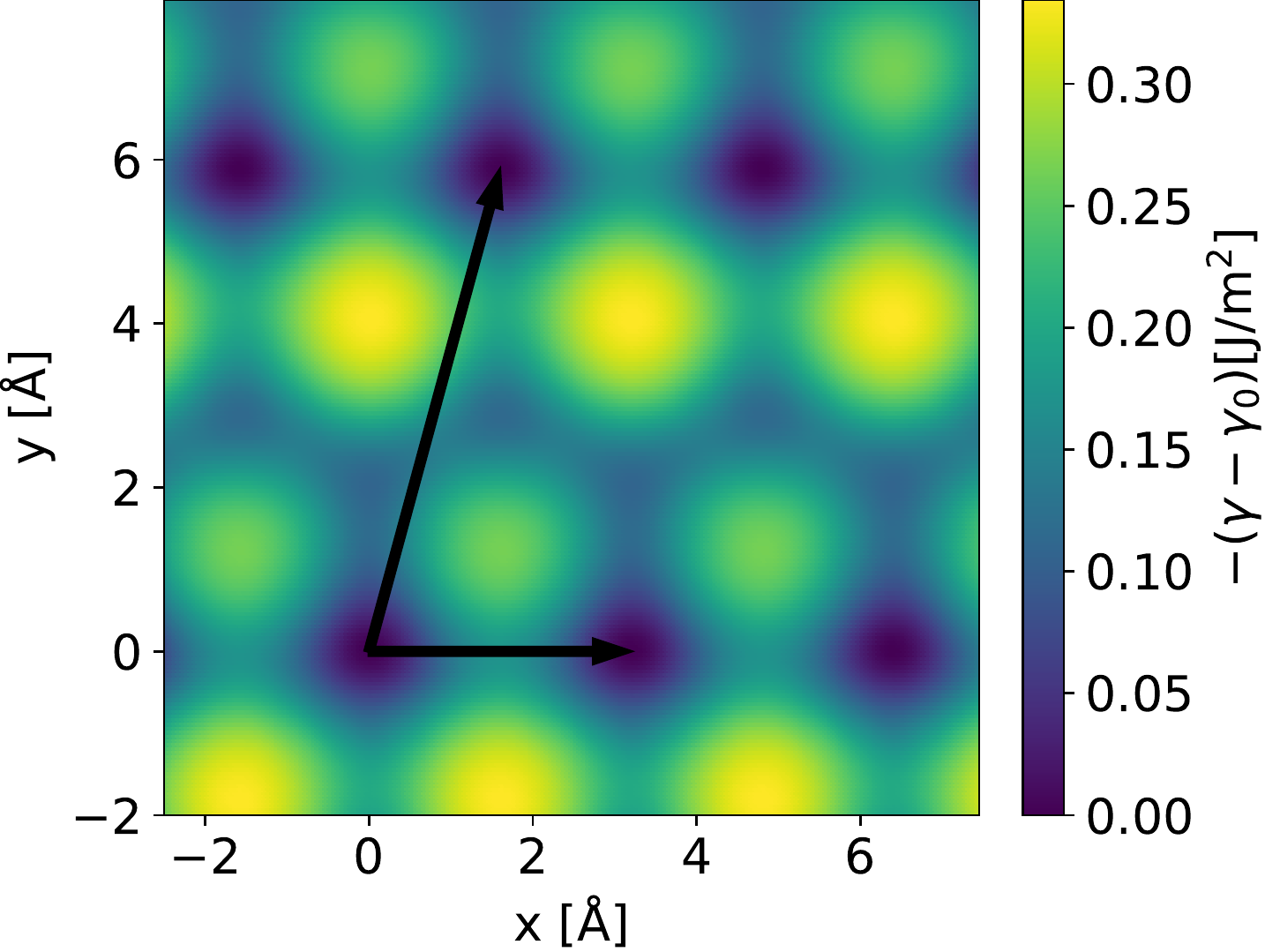}
    \label{fig:pyramidal.gammasurf}
}
\hfill
\subfloat[]{
    \includegraphics[width=0.9\linewidth]{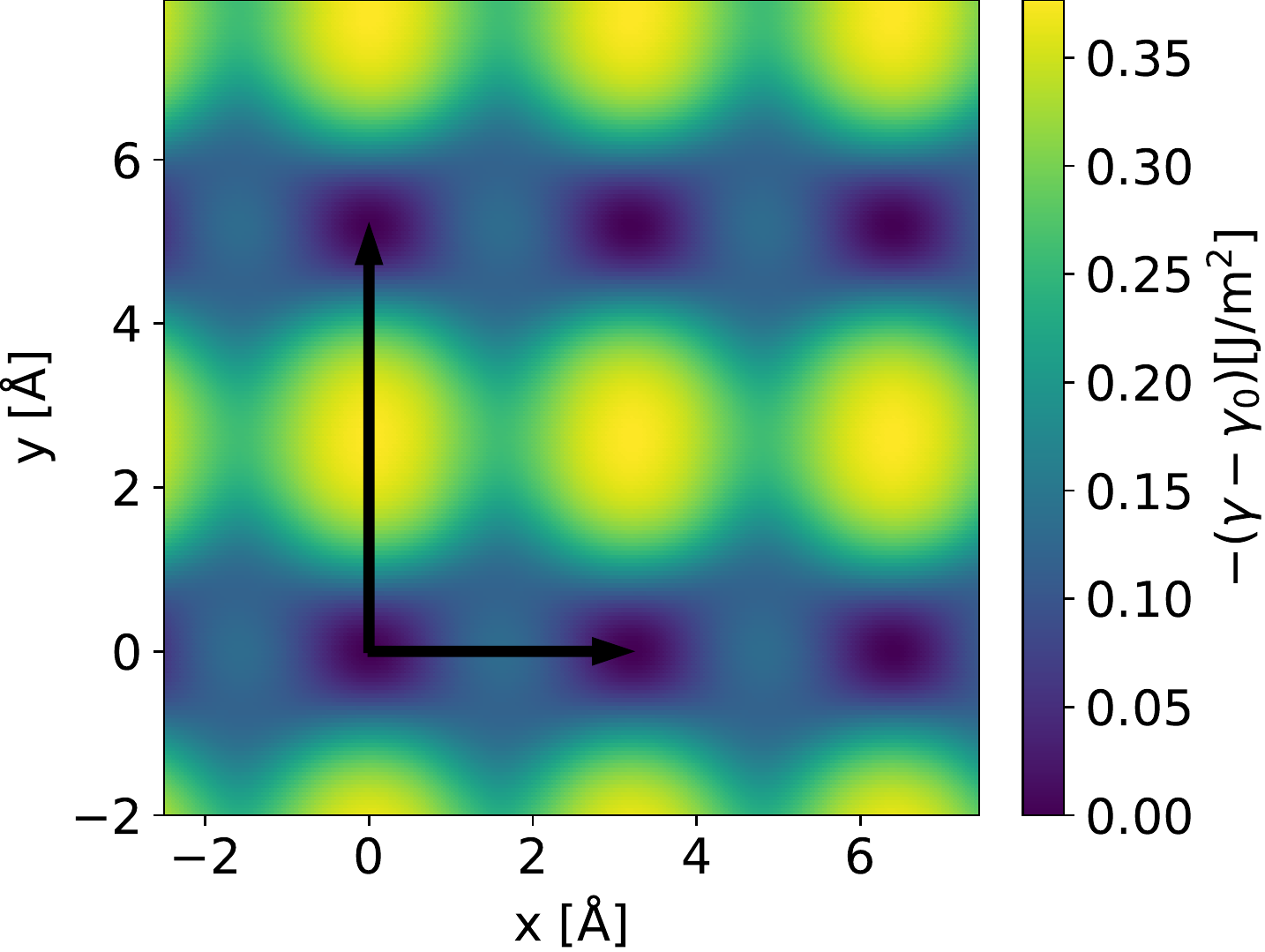}
    \label{fig:prismatic.gammasurf}
}
\caption{(a) Different slip planes of an hcp crystal. (b) The energy of hcp Mg as a function of glide parallel to the pyramidal plane evaluated at $d_0$ for each glide vector. (c) The energy of hcp Mg as a function of glide parallel to the prismatic plane evaluated at $d_0$.}
\label{fig:Mg_surfaces} 
\end{figure}

While slip in fcc predominantly occurs on (111) planes, there are often multiple slip planes in more complex crystal structures such as hcp. 
\Cref{fig:hcp_slip_planes} schematically shows two slip planes in hcp: basal slip, prismatic slip and pyramidal slip. 
Each plane has a different periodicity. 
\Cref{fig:pyramidal.gammasurf} and \cref{fig:prismatic.gammasurf} shows the energy at fixed spacing $d$ for the pyramidal and prismatic slip planes of hcp Mg. 
The DFT method to calculate these energy surfaces was the same as that used for Al in \Cref{fig:al_surfaces}, except that a plane wave cutoff of 650 eV was used. 

\subsection{Crystallographic models of twisted two-dimensional materials: triangular lattices and honeycombs}\label{sec:honey}

\mush{} facilitates the construction of crystallographic models of twisted two-dimensional materials. 
As described in \cref{sec:bicrystals}, most twist angles $\theta$ do not produce structures that have periodic boundary conditions. 
The imposition of periodic boundary conditions, therefore, requires an adjustment of the target twist angle $\theta$ by $\Delta \theta$ and some degree of strain within the twisted two-dimensional building blocks. 
We explore the variation in the error $\Delta \theta$ in the target twist angle and the strain within the twisted building blocks due to the imposition of periodic boundary conditions for a pair of triangular lattices.
This is of relevance for two-dimensional materials such as MoS$_2$ and graphene since their two-dimensional lattices are triangular. 
Of interest is the variation of $\Delta \theta$ and strain with twist angle $\theta$ and maximum super cell size. 

\mush{} determines the best super cell for a twisted bicrystal with target twist angle $\theta$ that is smaller than a user specified maximum.
The "best" super cell is defined as the super cell that minimizes  $\sqrt{\lambda_1^2+\lambda_2^2}$, where $\lambda_1$ and $\lambda_2$ are the non-zero eigenvalues of the strain matrix.
This strain metric is equal to zero for super cells that are commensurate with the lattices of the twisted bicrystal.  
The size of a super cell is measured in terms of the number of primitive two-dimensional unit cells of the fundamental slab building blocks. 

\begin{figure}
    \centering
    \subfloat[]{
    \includegraphics[width=\linewidth]{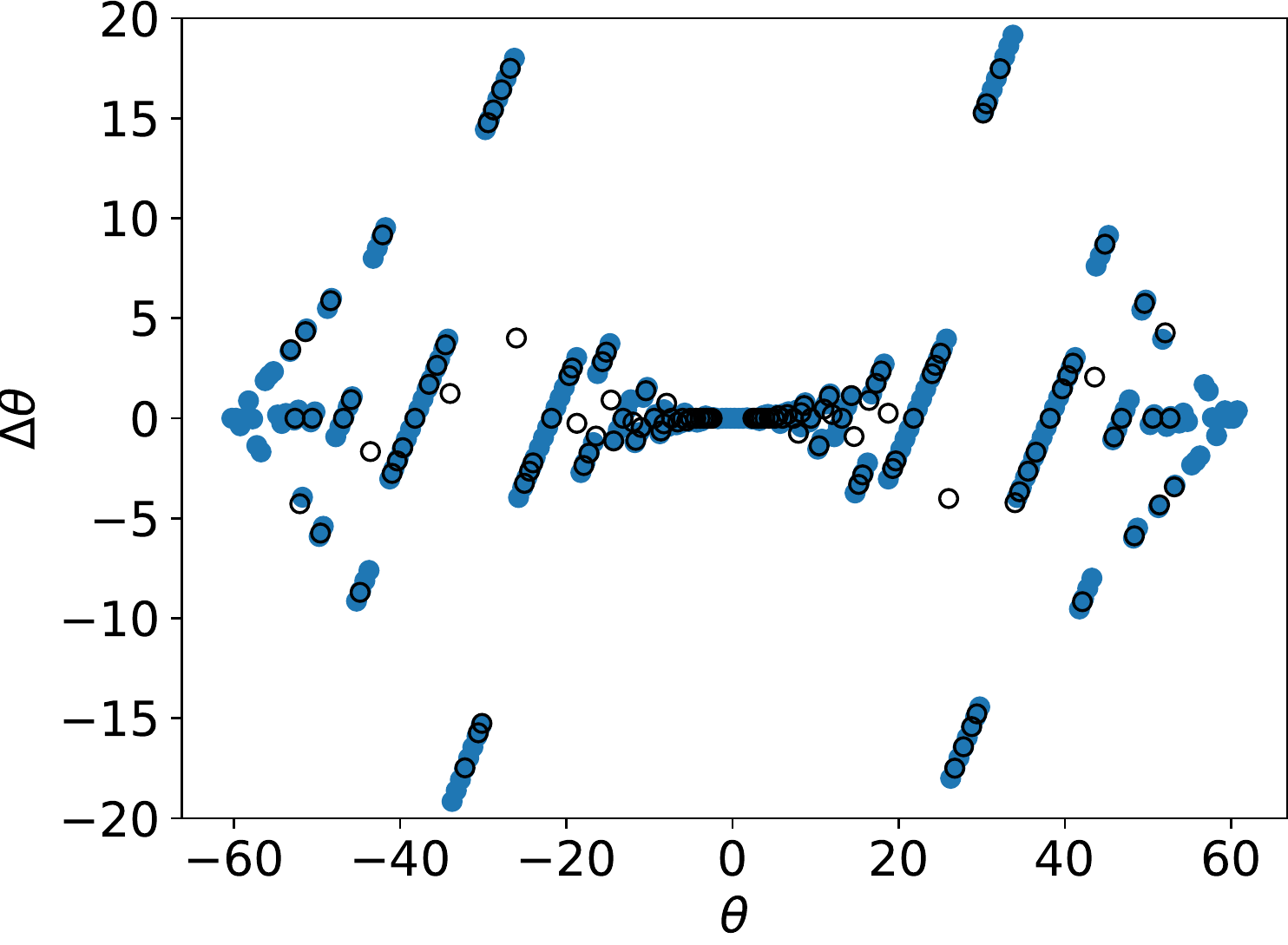}
    \label{fig:angleerror0}
}

\subfloat[]{
    \includegraphics[width=\linewidth]{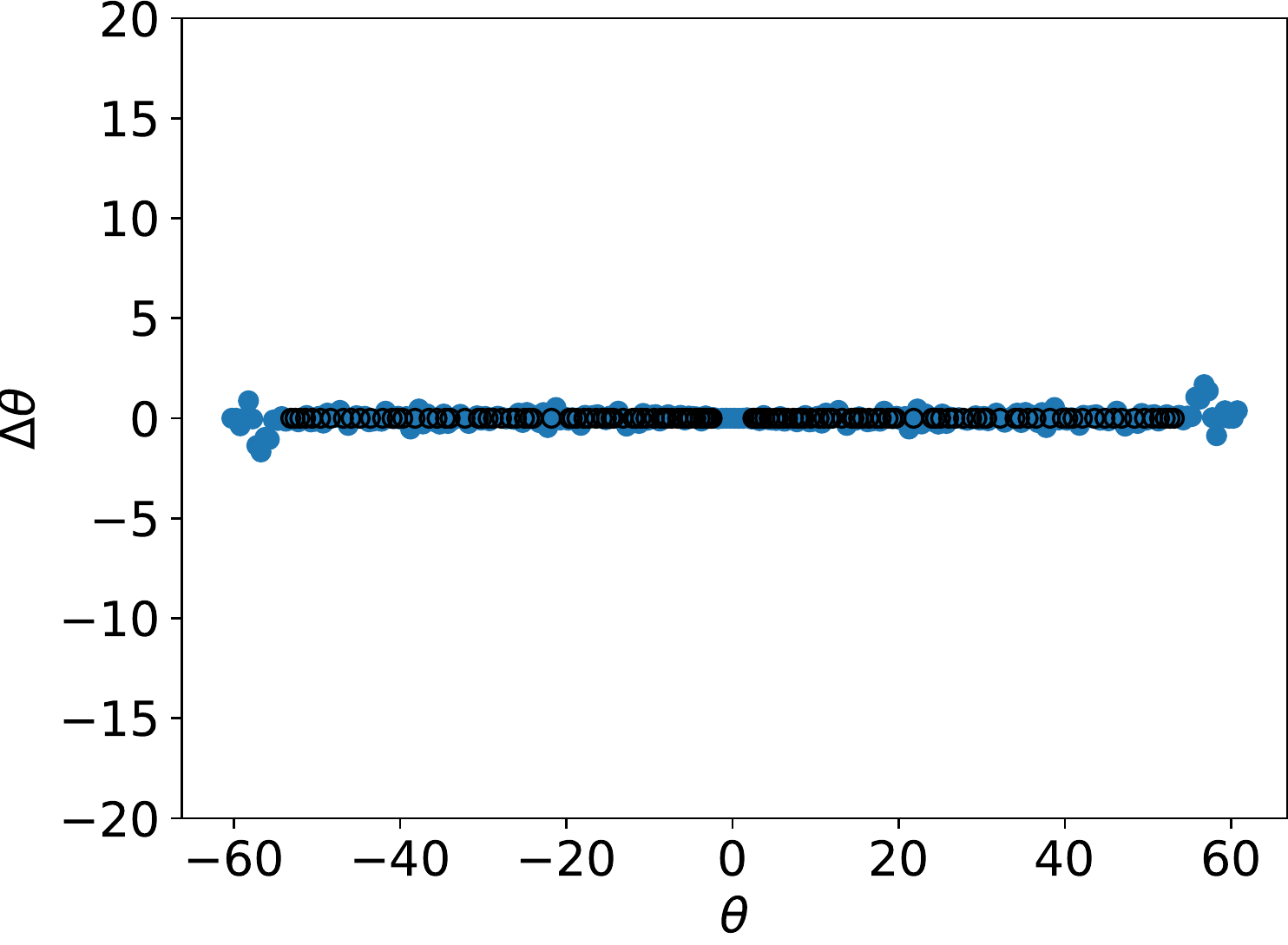}
    \label{fig:angleerror1000}
}
    \caption{(a) The error $\Delta \theta$ in the target rotation angle when using the Moir\'e lattice to determine the cell of the twisted bicrystal.
    (b) By considering {\it super cells} of the Moir\'e lattice to identify the cell of the twisted bicrystal, it is possible to dramatically reduce the error $\Delta \theta$. A cap of 1000 triangular lattice sites was imposed when enumerating super cells of the Moir\'e lattice. The black circles correspond to twist angles for which commensurate super cells exist.}
    \label{fig:degreeerror}
\end{figure}

\begin{figure}
    \centering
    \subfloat[]{
    \includegraphics[width=\linewidth]{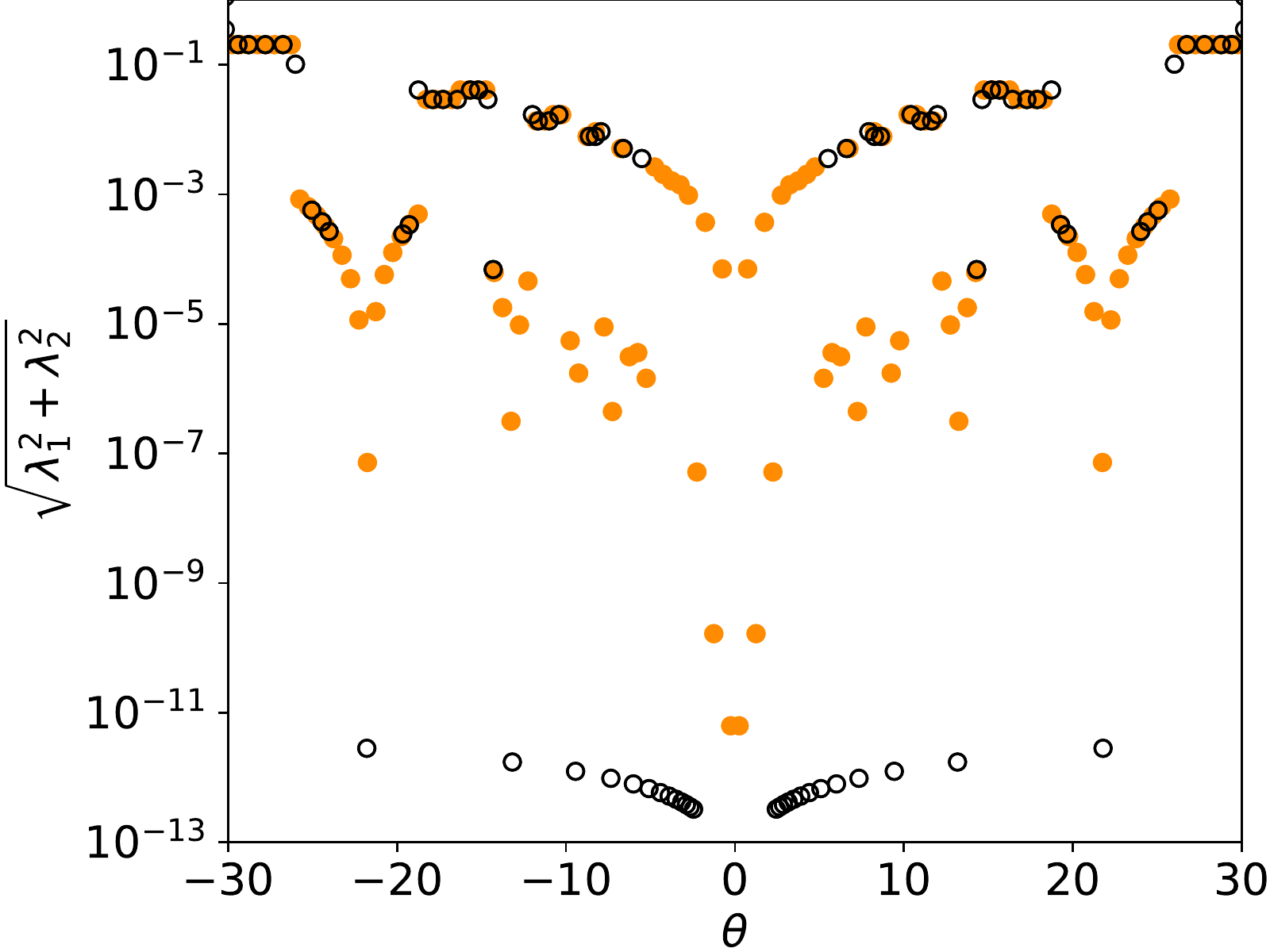}
    \label{fig:strainerror0}
}
\hfill
\subfloat[]{
    \includegraphics[width=\linewidth]{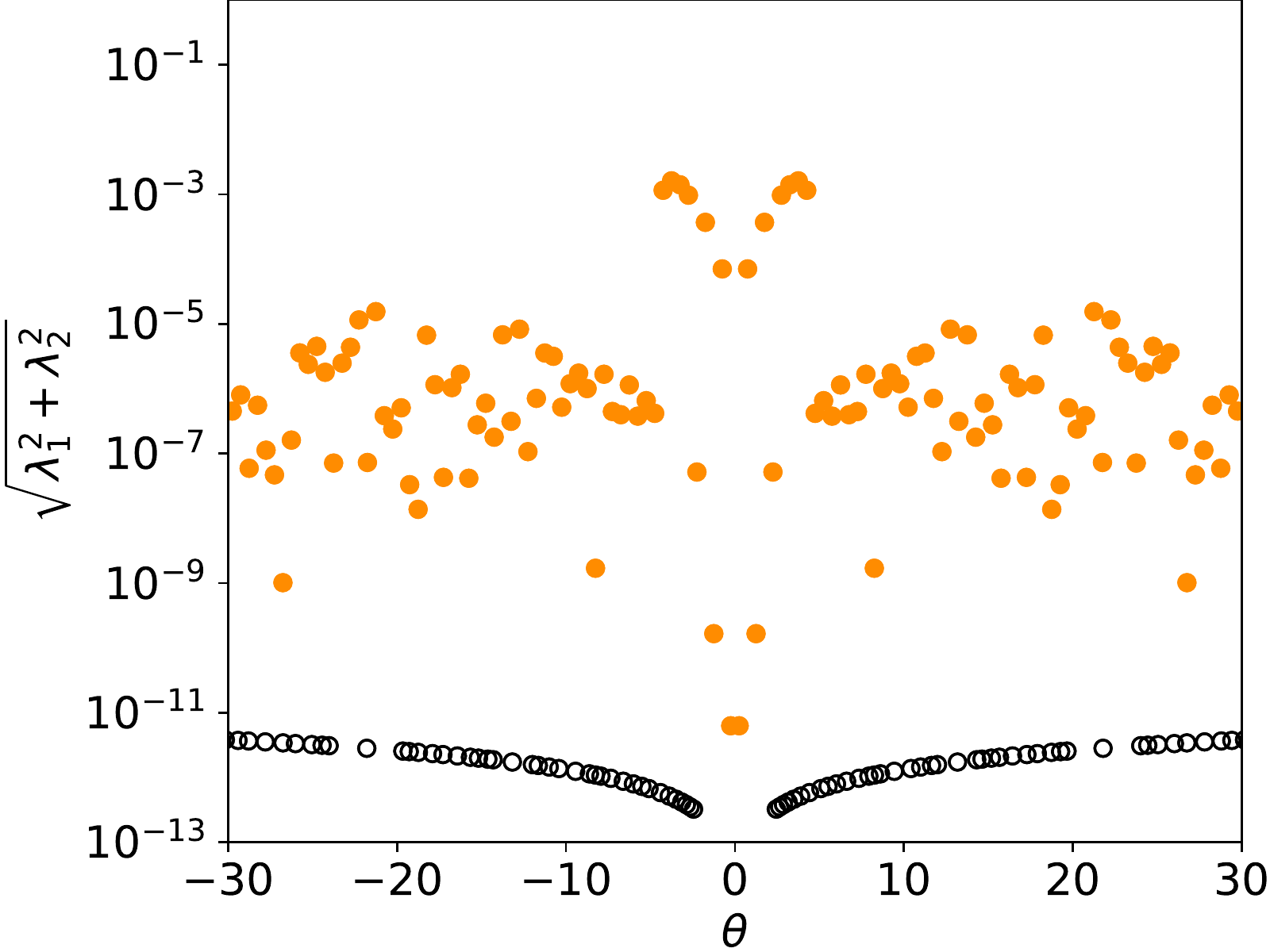}
    \label{fig:strainerror1000}
}
    \caption{(a) The strain (as measured with $\sqrt{\lambda_{1}^2+\lambda_{2}^2}$, where $\lambda_1$ and $\lambda_2$ are the non-zero eigenvalues of the strain matrix) as a function of the target twist angle $\theta$ when using the Moir\'e lattice to generate the cell of the twisted pair of triangular lattices. (b) The strain when using {\it super cells} of the Moir\'e lattice to generate the cell of the twisted triangular lattice (a cap of 1000 triangular lattice sites was imposed). The black circles correspond to twist angles for which commensurate super cells exist.}
    \label{fig:strainerror}
\end{figure}

\Cref{fig:degreeerror} and \Cref{fig:strainerror} shows $\Delta \theta$ and $\sqrt{\lambda_1^2+\lambda_2^2}$ as a function of $\theta$ for two scenarios. 
The error $\Delta \theta$ versus $\theta$ of \Cref{fig:angleerror0} is for super cells that were generated using the primitive Moir\'e lattice only. 
For small angles close to zero, the absolute error is small, however, for larger angles, the error $\Delta \theta$ can be very large. 
The black circles correspond to angles for which there is a commensurate super cell that can accommodate the twisted pair of bicrystals. 
As is evident in \Cref{fig:angleerror0}, the primitive Moir\'e lattice is not sufficiently large to identify the commensurate super cell for those special angles, as a large fraction of them have large errors $\Delta \theta$.
Similarly, the strain for super cells generated using the primitive Moir\'e lattice is also large as shown in \Cref{fig:strainerror0}.
\Cref{fig:angleerror1000} plots $\Delta \theta$ versus $\theta$ as determined by considering {\it super cells} of the Moir\'e lattice to identify the optimal super cell of the twisted bicrystals. 
A cap of 1000 primitive unit cells was used for \Cref{fig:angleerror1000} and \Cref{fig:strainerror1000}. 
The error in the target angle $\Delta \theta$ is dramatically reduced and almost all special angles for which commensurate super cells exist now have an error $\Delta \theta$ equal to zero, indicating that all commensurate super cells have been found. 

\section{Discussion and Conclusion}\label{sec:discuss}
Two-dimensional defects in bulk crystals play an out-sized role in determining the properties of many crystalline materials. 
The energy of two-dimensional extended defects are a crucial ingredient to a variety of meso-scale models of fracture and dislocations \cite{deshpande2002discrete, xie2006discrete,sills2013effect,koslowski2002phase,shen2003phase,shen2004incorporation,hunter2011influence,feng2018shearing,lu2000peierls,lu2005peierls}. 
The information of relevance for such models can be encapsulated in a generalized cohesive zone model that describes the energy of a pair of bicrystals as a function of the perpendicular separation and parallel glide of a pair of crystallographic planes. 
In this paper, we have described how such a generalized cohesive zone model can be formulated and how crystallographic models can be constructed for the first-principles electronic structure calculations that are needed to parameterize the cohesive zone model.
The \mush{} code automates the construction of periodic crystallographic models of decohesion and glide. 
It can also be used to generate crystallographic models for surface calculations. 
However, with the exception of the simplest materials, most surfaces undergo surface reconstructions to eliminate dangling bonds \cite{}. 
This requires an additional enumeration step \cite{thomas2010systematic}.

Twisted two-dimensional materials are currently attracting much attention due to the promise of emergent electronic properties in such structures \cite{carr2020electronic}. 
We have described an approach to construct crystallographic models of twisted bilayers within periodic unit cells. 
Only a subset of twist angles generate bilayer structures that are periodic. 
For all other angles, the imposition of periodic boundary conditions results in an error in the target twist angle and some degree of strain within the constituent bilayers. 
\mush{} constructs these crystallographic models and quantifies the twist angle error and the strain.

In the construction of cohesive zone models, a question often arises as to whether to allow for relaxations or not. 
When including relaxations, it is important to carefully extract the energy of homogeneous elastic relaxations of the adjacent slabs since the cohesive zone model should only describe the energy between the cleaved planes. 
These subtleties are discussed in great detail in \cite{enrique2017decohesion,van2004thermodynamics}.
For simple metals, empirical evidence suggests that relaxations do not need to be explicitly taken into considerations when parameterizing a cohesive zone model for decohesion \cite{enrique2017decohesion,van2004thermodynamics}.

Many applications require a generalized cohesive zone model for a multi-component solid. 
The cohesive zone model will then not only depend on the local concentration, but also on the local arrangement of different chemical species. 
This dependence can be accounted for with the cluster expansion approach \cite{sanchez1984generalized,van2018first} as has been done in the context of hydrogen embrittlement \cite{van2004thermodynamics} and fracture in Li-ion battery electrode materials \cite{qi2012chemically}.
Often cohesive zones must be treated as open systems to which mobile species can segregate and thereby alter cohesive properties. 
In these circumstances it is convenient to formulate cohesive zone models at constant chemical potential as opposed to constant concentration \cite{van2004thermodynamics,enrique2014solute,olsson2017intergranular}
A cluster expansion approach has also been used to describe the dependence of the $\gamma$ surface on composition and short-range ordering in a refractory alloys \cite{natarajan2020}. 

In summary, \mush{} is a code that automates the construction of crystallographic models within periodic unit cells to enable the construction of cohesive zone models and the study of twisted bilayers of two-dimensional materials with first-principles electronic structure methods. 
\mush{} can also be used to construct models with highly distorted local environments that are representative of dislocations, grain boundaries and surfaces for the purposes of training machine learned inter-atomic potentials. 

\section{Acknowledgement}

The scientific work in this paper was supported by the National Science Foundation DMREF grant: DMR-1729166 “DMREF/GOALI: Integrated Computational Framework for Designing Dynamically Controlled Alloy -Oxide Heterostructures”.
Software development was supported by the National Science Foundation, Grant No. OAC-1642433.
Computational resources provided by the National Energy Research Scientific Computing Center (NERSC), supported by the Office of Science and US Department of Energy under Contract No. DE-AC02-05CH11231, are gratefully acknowledged, in addition to support from the Center for Scientific Computing from the CNSI, MRL, and NSF MRSEC (No. DMR-1720256).

\section{Data Avaliability}
The data used as an example for the application of \mush{} cannot be shared at this time due to time limitations.

\bibliography{./references.bib}
\end{document}